\documentclass[a4paper,12pt]{article}%
\usepackage{amsmath,amssymb,amsthm}
\usepackage{natbib}
\usepackage{amsmath}
\usepackage{amsfonts}
\usepackage{setspace}
\usepackage{amssymb}
\usepackage{graphicx}%
\usepackage{color}
\usepackage{apptools}

\newtheorem{proposition}{Proposition}
\newtheorem{lemma}{Lemma}

\newtheorem{theorem}{Theorem}

\newtheorem{remark}{Remark}

\newtheorem{claim}{Claim}

\newcommand{\N}{\mathbb{N}}
\newcommand{\R}{\mathbb{R}}

\renewcommand{\d}{{\rm d}}

\newcommand{\calE}{\mathcal{E}}
\newcommand{\calM}{\mathcal{M}}

\DeclareMathOperator*{\argmax}{arg\,max}

\AtAppendix{\counterwithin{lemma}{section}}
\AtAppendix{\counterwithin{proposition}{section}}
\AtAppendix{\counterwithin{theorem}{section}}
\AtAppendix{\counterwithin{corollary}{section}}

\begin{document}

\title{Convex Cost of Information via Statistical Divergence\thanks{We are extremely grateful for insightful comments and suggestions to Alex Bloedel, Xiaoyu Cheng, Mark Dean, Tommaso Denti, Mira Frick, Yonggyun Kim, Xiaosheng Mu, Nathaniel Neligh, Pietro Ortoleva, Luciano Pomatto, Larry Samuelson, Philipp Strack, Omer Tamuz, Weijie Zhong, and various conference participants.
\newline  Bordoli: Yale University (davide.bordoli@yale.edu); Iijima: Princeton University (riijima@princeton.edu).}}

	\author{Davide Bordoli \and Ryota Iijima}
\date{\today}
\maketitle 

\begin{abstract}
This paper characterizes convex information costs using an axiomatic approach.
We employ mixture convexity and sub-additivity, which capture the idea that producing ``balanced" outputs is less costly than producing ``extreme” ones.
Our analysis leads to a novel class of cost functions that can be expressed in terms of Rényi divergences between signal distributions across states. This representation allows for deviations from the standard posterior-separable cost, thereby accommodating recent experimental evidence.
We also characterize two simpler special cases, which can be written as either the maximum or a convex transformation of posterior-separable costs.
\end{abstract}

\section{Introduction}
\subsection{Motivation and Overview}
In many economic environments, acquiring information is costly for a decision maker (DM). These costs can reflect a range of factors that are often difficult to observe, such as the resources the DM expends to gather data and the cognitive effort required to process it. An axiomatic approach offers a useful framework for organizing different types of information costs. Representation results, in particular, help identify which cost functions are consistent with desirable postulates, without requiring an explicit account of the underlying sources of the costs.

The main motivation of this paper is to axiomatically characterize {\it convex} cost functions over experiments. Convexity is a fundamental concept in producer theory and plays an important role in many areas of economics.\footnote{Beyond the current context, there have been other efforts to formalize convexity in non-standard economic domains; see, for example, \citet{richter2015} and \citet{murota2016}.} It captures the idea that producing ``balanced” outputs is less costly than producing ``extreme” ones. For instance, in a two-good economy, convexity implies that producing 1,000 units of each good is less costly than producing 2,000 units of only one good, assuming that the cost of producing 2,000 units is the same for either good. To formalize this idea in the context of information, the paper uses two distinct operations over experiments: mixture and bundling.

Our first main axiom is {\it mixture convexity}, which requires that the cost of a probabilistic mixture of two experiments is no greater than the expected cost of independently randomizing between them. This axiom reflects the idea that a mixture of experiments is more balanced than either individual experiment, thereby reducing the expected cost. The extreme case, {\it mixture linearity}, is satisfied by the standard approach based on posterior-separable cost functions.

Our second main axiom is {\it sub-additivity}, which states that acquiring a bundle of two independent experiments together should be less costly than acquiring them separately. This reflects the intuition that combining different experiments yields a more balanced acquisition of information.
To complete this idea, we also impose {\it identity additivity}, which requires that sub-additivity holds with equality when the two experiments are identical. In other words, combining identical experiments does not generate a balancing benefit and thus yields no cost reduction.

Our analysis extends  \citet{pomatto2023}, who characterize cost functions that satisfy mixture linearity and additivity—the extreme cases of mixture convexity and sub-additivity.
To illustrate their result, consider a finite state space $\Theta$ and signal space $S$, and let $\mu = (\mu_i)_{i \in \Theta}$ denote a generic experiment, where $\mu_i$ is the signal distribution conditional on state $i \in \Theta$.
They show that cost functions satisfying these axioms, along with Blackwell monotonicity and a form of continuity, can be represented by the {\it KL cost function} of the form:
\begin{equation}\label{eq:LLR}
C(\mu) = \sum_{i,j \in \Theta} \beta_{ij}  {\rm KL}(\mu_i \| \mu_j),
\end{equation}
where ${\rm KL}(\mu_i \| \mu_j) = \sum_s \mu_i(s) \log \frac{\mu_i(s)}{\mu_j(s)}$ is the {\it Kullback–Leibler (KL) divergence} between the signal distributions $\mu_i$ and $\mu_j$, and $(\beta_{ij})_{i,j \in \Theta}$ are nonnegative weights.\footnote{This cost function was originally referred to as the LLR cost; we adopt the current terminology for clarity.}
The KL divergence measures how informative an experiment is for distinguishing states $i$ and $j$, and the total cost aggregates these pairwise distinctions.

Our representation of convex cost functions builds on {\it Rényi divergence}, a generalization of KL divergence, along with its non-binary extension. Specifically, we use a parameterized family of divergences $D_{\alpha, \beta}(\mu)$, which includes both the weighted KL divergence in equation~(\ref{eq:LLR}) and the following form:
\begin{equation}\label{eq:D-alpha}
D_{\alpha,\beta}(\mu) = -\log \sum_{s \in S} \prod_{i \in \Theta} \mu_i(s)^{\alpha_i},
\end{equation}
for some non-unit vector  $\alpha \in [0,1]^\Theta$ satisfying $\sum_{i \in \Theta} \alpha_i = 1$. Like equation (\ref{eq:LLR}), this divergence quantifies the amount of information an experiment provides for the DM to distinguish among different states.
This form of divergence appears in statistical decision theory \citep[e.g.,][]{torgersen1991} and reduces to the standard Rényi divergence when the state space is binary.

Theorem~\ref{thm:subadditive} characterizes cost functions that satisfy mixture convexity, sub-additivity, identity additivity, and Blackwell monotonicity. Building on the techniques developed in \cite{mu2021} and \cite{farooq2024}, we show that such cost functions can be represented by the {\it Max-Rényi cost function} of the form
$$
C(\mu) = \max_{m \in M}\int  D_{{\alpha,\beta}}(\mu) \, d m(\alpha, \beta),
$$
where $M$ is a set of measures $m$ over divergence parameters $(\alpha, \beta)$.
 This generalizes the KL cost function (\ref{eq:LLR}) in two ways. First, the cost is determined by  Rényi divergences (beyond KL divergences) among signal distributions across states. 
Second, the representation involves the maximum operator over a set \(M\) of measures over divergences, so that each experiment is evaluated using the most expensive one.

As we clarify in Proposition~\ref{prop:additive-linear}, the maximum over measures in $M$ is crucial to allow for strict sub-additivity; the cost function becomes additive when $M$ is a singleton.
In contrast, strict mixture convexity can be accommodated through two distinct channels: (i) using the maximum operator, i.e., $|M| > 1$, and/or (ii) using Rényi divergences of the form (\ref{eq:D-alpha}) instead of KL divergences.

Theorem~\ref{prop:d-linear} characterizes two simpler special cases of the Max-Rényi cost function that isolate the two channels described above. First, we consider the {\it Max-KL cost function} of the form
$$
C(\mu) = \max_{\beta \in B} \sum_{i,j \in \Theta} \beta_{ij} \, {\rm KL}(\mu_i\| \mu_j),
$$
for some set $B$ of coefficients. This form was suggested by \cite{pomatto2023} as a ``possible definition of convex cost functions over experiments.” We show that this corresponds to the special case of the Max-Rényi cost that satisfies {\it dilution linearity}, i.e., mixture linearity holds when an experiment is mixed with an uninformative experiment.
Second, as an alternative special case that does not invoke the maximum operator, we consider the {\it Rényi cost function}, i.e., the cost is either proportional to (\ref{eq:D-alpha}) or a KL cost. This cost function has the advantage of being differentiable with respect to signal probabilities. We show that it corresponds to the special case that satisfies {\it independence}, which is an ordinal version of mixture linearity.

As shown by \citet{pomatto2023}, KL cost functions are posterior-separable. In contrast, Max-Rényi cost functions allow for departures from posterior separability. Relative to the general representation, the two special cases above maintain disciplined forms of this departure.
Specifically, Max-KL cost functions correspond exactly to the class of Max-Rényi cost functions that can be written as the maximum of posterior-separable costs. Likewise, Rényi cost functions correspond precisely to the special case that can be written as convex monotone transformations of posterior-separable costs.

Section~\ref{sec:application} applies the Max-Rényi cost to optimal information acquisition problems, highlighting the role of mixture convexity. It is well known that under posterior-separable cost functions, which are mixture linear,  DM's optimal policy can be achieved using at most $|\Theta|$ distinct actions. This prediction may seem somewhat extreme, particularly when DM’s action set is large. To confirm this intuition, we examined the lab experiments conducted by \cite{dean2023}, which involve two states and three actions, and observed that a significant proportion of subjects selected all three actions with positive frequencies. We show that a simple special case of the Rényi cost can account for such choice patterns.

\subsection{Related Literature}

This paper contributes to the literature on the axiomatic foundations of information cost. Several studies characterize cost functions from a revealed-preference perspective, taking DM's choice data as primitive \citep[e.g.,][]{caplin2015, matvejka2015, de2017, ellis2018, denti2022, caplin2022, mensch2023}. Most of these studies focus on characterizing cost functions that are either mixture linear or fully general. Relatedly, these papers demonstrate that the mixture convexity of cost functions is not a testable property within their framework.\footnote{This is somewhat analogous to the classical finding that the convexity of preferences is not testable in the Afriat setting. } In this paper, we adopt the approach of \cite{pomatto2023}, which takes cost functions over experiments as primitive and allows us to trace out the implications of mixture convexity.

Posterior-separable cost functions, and their special case of the Shannon entropy cost \citep{sims2003}, are frequently used in applications, partly due to their tractability \citep[see, e.g.,][for a survey]{mackowiak2023}. At the same time, several papers point to certain limitations of posterior-separable costs. One notable issue is that these cost functions depend on DM's prior belief, which can lead to conceptual and practical challenges \citep[e.g.,][]{gentzkow2014, mensch2018, denti2022b}. This paper shares a similar perspective to \cite{denti2022}, who highlights the role of mixture linearity. He provides a revealed-preference characterization of posterior-separable costs, which is shown to be inconsistent with the dataset in \cite{dean2023}.
As a solution, he suggests using convex transformations of posterior-separable costs. Our paper provides an axiomatic foundation for a particular functional form (Rényi cost) within this class.

A strand of literature interprets the cost of information as an indirect cost arising from dynamic experimentation, where  DM is endowed with an exogenous direct cost of information \citep[e.g.,][]{morris2019, denti2022b, hebert2023}. \cite{bloedel2021} provide a general characterization of such cost functions, which are shown to satisfy the sequential learning-proof property.
As we discuss in Section~\ref{sec:discussion}, several of our main axioms can be justified from this perspective.

Several papers characterize solutions to DM's optimal information acquisition problems \citep[e.g.,][]{matvejka2015, steiner2017, caplin2019, caplin2022}.  Most of them focus on posterior-separable costs, with a recent exception by \cite{bloedel2025}, who use a representation based on a non-binary divergence notion called $f$-informativeness.

Beyond the context of information cost, quantitative measures of experiments have been studied in various economic settings \citep[e.g.,][]{cabrales2013, frankel2019}.
As we discuss in Appendix~\ref{app:others}, our representations can be related to the measures of information used in learning \citep{moscarini2002, frick2022} and privacy \citep{dwork2006}.

\section{Setting}

Let $\Theta$ denote a finite set of states with $|\Theta| \geq 2$. An \textbf{\textit{experiment}} $\mu$ consists of (i) a space of signals $S$, endowed with some $\sigma$-finite measure $\lambda$, and (ii) a probability measure $\mu_i$ over signals for each state $i \in \Theta$ that is absolutely continuous with respect to $\lambda$. For technical reasons, we focus on experiments that are bounded; that is, for each distinct pair of states $i, j \in \Theta$, the log likelihood ratio $\log \frac{d\mu_i}{d\mu_j}(s)$ is bounded almost surely with respect to $\lambda$.  Let $\mathcal{E}$ denote the class of all such experiments.

DM is endowed with an \textbf{\textit{information cost function}}, a mapping $C: \mathcal{E} \to \mathbb{R}_+$ that specifies the cost $C(\mu)$ required to acquire each experiment $\mu$.
In this paper, we remain agnostic about the origin of the information cost $C$. Our aim is to understand which representations of $C$ can be characterized by reasonable postulates on the information cost.\footnote{Our analysis does not rule out cost functions that depend on DM's beliefs, such as posterior-separable costs. Unless necessary, we do not explicitly denote the dependence of $C$ on beliefs.}



The key ingredient of our analysis is divergence functions. 
We build on  \textbf{\textit{Rényi divergence}}, which is a common tool to quantify the difference between a  pair of signal distributions. Specifically, the divergence between $\mu_i$ and $\mu_j$ is given by
\[
R_t(\mu_i \|\mu_j) :=  
\frac{1}{t-1}\log \int \left(\frac{d\mu_i(s)}{d\lambda(s)}\right)^{t}\left(\frac{d\mu_j(s)}{d\lambda(s)}\right)^{1-t}  d\lambda(s)
\]
parameterized by $t\in [1/2, 1)$.\footnote{The divergence can be defined for $t>1$, but this case will be ruled out by our mixture-convexity axiom.} 
Rényi divergence nests KL divergence as an extreme case as $t\to 1$, i.e.,  
 \[
R_1(\mu_i \| \mu_j):=\lim_{t\to 1}R_t(\mu_i \| \mu_j) =  \int \log \frac{d\mu_i}{d\mu_j}(s) d\mu_i(s) =: {\rm KL}(\mu_i\| \mu_j).
\]
Rényi divergence is a standard tool in statistics and econometrics \citep[see, e.g.,][]{van2014}.  
\cite{mu2021} show that Rényi divergence can be used to compare the value of experiments in binary-state problems.


Our paper employs a non-binary extension of Rényi divergence that quantifies the differences among signal distributions across states: 
\[
D_\alpha (\mu):=  \frac{1}{\alpha_{\rm max} -1} \log \; \int \prod_{i\in\Theta} \left(\frac{d\mu_i(s)}{d\lambda(s)}\right)^{\alpha_i}d \lambda(s) 
\] 
where $\alpha\in\mathbb R^\Theta_+\setminus\{e_i:i\in\Theta\}$ is such that $\sum_{i\in\Theta}\alpha_i=1$, $\alpha_{\rm max}:=\max_{i}\alpha_i$, and $e_i$ denotes the $i$'th unit vector.\footnote{When signals are finite, the divergence takes the form 
\[
D_\alpha (\mu)=  \frac{1}{\alpha_{\rm max}  -1} \log \; \sum_{s\in S} \prod_{i\in\Theta}  \mu_i (s)^{\alpha_i}.
\]
}
Observe that 
$D_\alpha(\mu)=0$ whenever $\mu_i$ is independent of $i$. 
This extension of Rényi divergence was proposed by \cite{toussaint1974} and used in statistics to evaluate the performance of experiments in estimation and hypothesis testing.\footnote{See \cite{pardo2018} for a textbook treatment. It is also known as Hellinger transform and used in statistical decision theory \citep[e.g.,][]{torgersen1991}.}

The limit of $D_{\alpha}$ as $\alpha\to e_i$ corresponds to weighted KL divergences. Specifically, we have 
\begin{equation}\label{eq:limit-KL}
\lim_{\gamma\nearrow 1} D_{\gamma e_i+(1-\gamma)\beta}=\sum_{j\not=i}\beta_j{\rm KL}(\mu_i \| \mu_j)
\end{equation}
for any $i\in\Theta$ and $\beta \in \mathbb R^\Theta_+$ such that $ \sum_{j\in\Theta}\beta_j=1$.


%




\section{Main Results}


 \subsection{Main Axioms}\label{sec:sub-additive}

This section introduces our main axioms that capture the basic idea of convex information cost: it is more costly to produce extreme information than balanced information.
We make use of two operators over experiments to formalize these axioms.

\

\noindent{\bf Mixture of experiments}: Given experiments $\mu = (S, (\mu_i)_{i \in \Theta})$ and $\nu = (T, (\nu_i)_{i \in \Theta})$ with $S \cap T = \emptyset$ and $\alpha \in (0, 1)$, their mixture is defined as
$$
\alpha \mu + (1-\alpha) \nu := (S \cup T, (\alpha \mu_i + (1-\alpha) \nu_i)_{i \in \Theta}),
$$
i.e., a signal is drawn from $\mu$ with probability $\alpha$, and from $\nu$ with the remaining probability $1-\alpha$.

\

Our main focus is on cost functions that are \textit{\textbf{mixture convex}}:
$$
C(\alpha \mu + (1-\alpha) \nu) \leq \alpha C(\mu) + (1-\alpha) C(\nu)
$$
for all $\mu, \nu$ and $\alpha \in (0, 1)$. Thus, the cost of a probabilistic mixture of two experiments is less than the expected cost of independently randomizing between them. As noted in Introduction, this reflects the idea that probabilistic mixtures provide more balanced information than the original experiments.

Most cost functions used in applications take the extreme form of being \textit{\textbf{mixture linear}}:
$$
C(\alpha\mu + (1-\alpha)\nu) = \alpha C(\mu) + (1-\alpha)C(\nu)
$$
for all $\mu, \nu$ and $\alpha \in (0, 1)$. Mixture linear cost functions include the KL cost and, more generally, posterior-separable costs. While mixture linear cost functions are tractable, the next section will demonstrate that their predictions can conflict with empirical findings.

\

\noindent{\bf Bundle of experiments}: Given two experiments $\mu=(S, (\mu_i)_{i\in\Theta})$ and $\nu=(T, (\nu_i)_{i\in\Theta})$, their product is defined by
$$
\mu\otimes \nu:=(S\times T, (\mu_i\times \nu_i)_{i\in\Theta}),
$$
where signals are drawn independently from both experiments $\mu$ and $\nu$, and both are observed.

\

Our main postulate is that cost functions are \textit{\textbf{sub-additive}}:
$$
C(\mu \otimes \nu) \leq C(\mu) + C(\nu)
$$
for all $\mu, \nu$.
Thus, acquiring different experiments together costs less than the sum of the individual costs.
The basic idea is that bundling two different experiments provides more balanced information acquisition than the individual experiments.
However, we need an additional axiom to complete this idea. That is, we impose additivity whenever identical experiments are bundled, in which case the balancing effect is absent. Denote by $\mu^{\otimes 2} := \mu \otimes \mu$ the product of identical experiments, and inductively, $\mu^{\otimes k} := \mu^{\otimes (k-1)} \otimes \mu$ for every $k \in \mathbb{N}$. Thus, under $\mu^{\otimes k}$, DM observes $k$ i.i.d. signal draws from $\mu$.
The cost function $C$ is \textit{\textbf{identity additive}} if
$$
C(\mu^{\otimes k}) = kC(\mu)
$$
for all $\mu$ and $k \in \mathbb{N}$.
The combination of the two axioms ensures
$$
C(\mu\otimes \nu)\leq C(\mu^{\otimes 2})=C(\nu^{\otimes 2})
$$
under any experiments with $C(\mu)=C(\nu)$, i.e., the more balanced experiment $\mu\otimes \nu$ is cheaper than the extreme experiments $\mu^{\otimes 2}$ and $\nu^{\otimes 2}$.\footnote{Identity additivity can also be interpreted as the canonical assumption in sequential learning models, dating back to \cite{wald1945}, where the cost is proportional to the number of signal draws.}

Finally, recall that the KL cost function is \textit{\textbf{additive}}, i.e., for all $\mu, \nu$,
$$
C(\mu\otimes \nu)= C(\mu)+C(\nu).
$$

\subsection{General Representation}

To describe representations, we introduce some notations. Let $A_+ := \{\alpha \in \mathbb R^\Theta_+: \sum_{i \in \Theta} \alpha_i = 1\} \setminus \{e_i: i \in \Theta\}$ denote the parameter space of  Rényi divergence $D_\alpha$.  
As observed in (\ref{eq:limit-KL}), $\lim_{n \to \infty} D_{\alpha_n}$ corresponds to weighted KL divergences $\sum_{j \ne i} \beta_j {\rm KL}(\mu_i \| \mu_j)$ when $\alpha_n - e_i = \frac{1}{n}(\beta - e_i)$. 
Given this, we unify Rényi divergence and KL divergence by writing
\[
D_{\alpha, \beta}(\mu) :=
\begin{cases}
D_\alpha(\mu) & \text{if } \alpha \in A_+, \\
\sum_{j \ne i} \beta_j {\rm KL}(\mu_i \| \mu_j) & \text{if } \alpha = e_i.
\end{cases}
\]
Let \({\mathcal M}\) denote the space of Borel measures on the expanded parameter space $\{(\alpha, \beta) \in \mathbb R^\Theta_+ \times \mathbb R^\Theta_+: \sum_{i \in \Theta} \alpha_i = \sum_{i \in \Theta} \beta_i = 1\}$, endowed with the weak-convergence topology. 

The cost function is \textit{\textbf{Blackwell monotone}} if $C(\mu) \geq C(\nu)$ whenever \(\mu\) Blackwell-dominates \(\nu\), i.e., if there exists a measurable map \(\psi: S \to \Delta(T)\) such that \(\nu_i(E) = \int \psi_s(E) \, d\mu_i(s)\) for each \(i \in \Theta\) and measurable \(E \subseteq T\). 

\begin{theorem}\label{thm:subadditive}  
\(C\) is Blackwell monotone, mixture convex, sub-additive, and identity additive if and only if there exists a compact set \(M \subseteq {\mathcal M}\) such that for each \(\mu\),
\[
C(\mu) = \max_{m \in M} \int D_{{\alpha,\beta}}(\mu) \, dm(\alpha, \beta).
\]  
\end{theorem}

We refer to this representation as a \textbf{\textit{Max-Rényi cost function}}. This generalizes the KL cost function in two ways. First, the cost is determined by extended Rényi divergences (beyond KL divergences) among signal distributions across states. 
Here, each divergence $D_{\alpha,\beta}$ quantifies the usefulness of experiment $\mu$ in distinguishing among states; each divergence need not be symmetric across states, allowing some states to be harder to distinguish than others.\footnote{This flexibility is crucial in applications and is inconsistent with the Shannon entropy cost. See, e.g., \cite{hebert2021, morris2022, dean2023}.}  
Second, the representation involves the maximum operator over a set \(M\) of measures over divergences. Specifically, each experiment is evaluated at the measure that makes it the most expensive. 

To prove Theorem~\ref{thm:subadditive}, in the Appendix we first characterize a more general representation that allows for divergences with non-positive $\alpha$. To this end, we build on the matrix majorization result in \cite{farooq2024}, which is an extension of \cite{mu2021} to a non-binary state space.  
They show that if $D_{\alpha, \beta}(\mu) > D_{\alpha, \beta}(\nu)$ for all parameters $\alpha, \beta$ (allowing for non-positive $\alpha$), then \(\mu^{\otimes k}\) Blackwell-dominates \(\nu^{\otimes k}\) for all sufficiently large \(k\). Based on this result and our axioms, we show that \(C\) can be expressed as a homogeneous, monotone, and sub-additive functional of divergences, which is then represented as the maximum of integrations over divergences. Finally, we use mixture convexity to rule out measures supported on divergences with non-positive $\alpha$.

\subsection{Special Cases}

This section introduces two simpler special cases of our representation: Max-KL and Rényi cost functions.  
As a starting point, we clarify how each ingredient of the general representation contributes to relaxing additivity and mixture linearity. To this end, the following proposition characterizes the implications of imposing additivity and mixture linearity, respectively. 

\begin{proposition}\label{prop:additive-linear}
Take $C$ that is Blackwell monotone, mixture convex, sub-additive, and identity additive. Then the following holds:
\begin{enumerate}
\item $C$ is additive if and only if \(C\) is represented by a Max-Rényi cost function with \(|M| = 1\).  
\item $C$ is mixture linear if and only if \(C\) is represented by a KL cost function. 
\end{enumerate}
\end{proposition}

The first part shows that imposing additivity corresponds to the case where \(M\) is a singleton. This reflects the fact that both KL divergence and Rényi divergence are additive, i.e., $D_{{\alpha,\beta}}(\mu \otimes \nu) = D_{{\alpha,\beta}}(\mu) + D_{{\alpha,\beta}}(\nu)$. Therefore, the maximum operator in the general representation is essential for capturing strict sub-additivity.  
This result can be viewed as a many-state extension of Theorem 2 in \cite{mu2021}, which characterizes Blackwell monotone and additive functions over experiments.\footnote{Relatedly, \cite{mu2024} characterize monotone and additive functions defined on one-dimensional random variables.}

The second part shows that imposing mixture linearity reduces the representation to the KL cost function, reflecting the fact that KL divergence is mixture linear while Rényi divergences are not. Moreover, it shows that mixture linearity implies additivity (under the other axioms).  
This result suggests that Max-Rényi cost functions can accommodate strict mixture convexity through two distinct channels:  
(i) the maximum operator, i.e., $|M| > 1$, and  
(ii) Rényi divergences that are not KL divergences, i.e., $\alpha \in A_+$.

A special case that isolates the first channel is the \textbf{\textit{Max-KL cost function}}: for some compact set \(B \subseteq \mathbb{R}_{+}^{\Theta \times \Theta}\),
\begin{equation}\label{eq:maxKL}
C(\mu) = \max_{\beta \in B} \sum_{i,j \in \Theta} \beta_{ij} \, {\rm KL}(\mu_i \| \mu_j).
\end{equation}
This representation was originally proposed by \cite{pomatto2023} as a ``possible definition of convex cost.”

An alternative special case that isolates the second channel is the \textbf{\textit{Rényi cost function}}, where \(M\) is a singleton concentrated on a single (extended) Rényi divergence. That is, the cost is given by either 
\begin{equation}\label{eq:pure R}
C(\mu) = \lambda D_{\alpha, \beta}(\mu).
\end{equation}
for some \((\alpha, \beta)\) with \(\alpha \in A_+\) and \(\lambda \geq 0\), or  a KL cost.  Unlike the Max-KL cost, this representation always satisfies additivity, and has the advantage of being differentiable with respect to signal probabilities.

The following theorem characterizes these two special cases by imposing weaker forms of mixture linearity.\footnote{To prove the first part of the theorem, the assumption of mixture convexity is not required.}

\begin{theorem}\label{prop:d-linear}
Let \(C\) be Blackwell monotone, mixture convex, sub-additive, and identity additive. Then the following holds:
\begin{enumerate}
\item \(C\) is represented by a Max-KL cost function if and only if \(C\) is \textbf{\textit{dilution linear}}, i.e.,  
\[
C(\alpha \mu + (1 - \alpha) \phi) = \alpha C(\mu) + (1 - \alpha) C(\phi)
\]
for all \(\mu, \phi\) and \(\alpha \in (0,1)\), whenever \(\phi\) is uninformative.

\item \(C\) is represented by a Rényi  cost function if and only if \(C\) satisfies \textbf{\textit{independence}}, i.e., 
\[
C(\alpha \mu + (1 - \alpha) \nu) \geq C(\alpha \mu' + (1 - \alpha) \nu) \iff C(\mu) \geq C(\mu')
\]
for all \(\mu, \mu', \nu\) and \(\alpha \in (0,1)\).
\end{enumerate}
\end{theorem}

The first part shows that Max-KL cost functions can be obtained by imposing linearity on mixtures with uninformative experiments. One way to interpret this axiom is to view a mixture \(\alpha \mu + (1 - \alpha)\phi\), where \(\phi\) is uninformative, as providing no balancing effect relative to the original experiment \(\mu\), thereby motivating linearity across such mixtures.\footnote{The analogue of this condition for the product operation is \(C(\mu \otimes \nu) = C(\mu) + C(\phi)\) whenever \(\phi\) is uninformative. This is guaranteed by identity additivity and Blackwell monotonicity.}

The second part characterizes Rényi cost functions by the independence axiom, which is an ordinal version of mixture linearity.\footnote{This result is structurally similar to Proposition F.1 in \cite{mu2024}, which characterizes cumulant generating functions as monotone and additive functions of one-dimensional random variables satisfying the independence axiom. However, the domain and proof arguments differ from ours.} While this axiom allows for violations of mixture linearity, it imposes a consistency requirement analogous to the independence axiom in the context of risk preferences. For example, the axiom has a flavor of dynamic consistency: if we interpret mixtures as two-stage procedures, the axiom guarantees that DM's comparison between \(\mu\) and \(\mu'\) after the resolution of the first stage remains consistent with the comparison between \(\alpha \mu + (1 - \alpha)\nu\) and \(\alpha \mu' + (1 - \alpha)\nu\).

\begin{remark}\rm \
\begin{enumerate}
\item In contrast to Proposition~\ref{prop:additive-linear}, KL cost functions can also be characterized by Blackwell monotonicity, additivity, and dilution linearity (see Appendix~\ref{app:KL2}). This characterization is analogous to Theorem 1 in \cite{pomatto2023}, except that they impose a continuity axiom (which they conjecture to be unnecessary). One technical difference is that we focus only on bounded experiments, while they allow for a broader class of experiments.

\item One advantage of the Rényi cost (that is not mixture linear) is that it can accommodate signals that perfectly reveal states, unless such signals occur with probability one in every state. These types of experiments are often used in applications such as bandit problems. The cost of these experiments is unbounded under the Max-KL cost.
\end{enumerate}
\end{remark}

\subsection{Connections to Posterior-separable Costs}\label{sec:PS}
We relate our representations to the standard approach in the literature that uses \textit{\textbf{posterior-separable}} cost functions. Fix a DM's prior belief \( q \in \Delta(\Theta) \) over states. Such cost functions take the form
\[
C(\mu) = \int H(p) \, d\pi^\mu(p) - H(q),
\]
where \( H: \Delta(\Theta) \to \overline{\mathbb{R}} \) is a convex function, and \( \pi^\mu \in \Delta(\Delta(\Theta)) \) is the distribution of DM’s posterior belief induced by experiment \( \mu \). The functional form of \( H \) may depend on the prior \( q \); if not, the cost is said to be uniformly posterior separable.

As shown by \cite{pomatto2023}, KL cost functions are posterior separable, with 
\[
H(p) = \sum_{i,j \in \Theta} \beta_{ij} \frac{p_i}{q_i} \log\left(\frac{q_j}{p_j}\right).
\]
Any posterior-separable cost function is mixture linear.\footnote{\cite{mensch2018} further shows that any mixture-linear cost can be expressed in this form under certain regularity conditions.}
Therefore, by the second part of Proposition~\ref{prop:additive-linear}, the intersection of Max-Rényi costs and posterior-separable costs consists only of KL cost functions.

Two principled extensions of posterior-separable cost functions have been proposed in the literature.
One approach is to take a collection \( \mathcal{H} \) of functions \( H \), and define the cost as the maximum over these, i.e.,
\begin{equation}\label{eq:max-PS}
C(\mu) = \max_{H \in \mathcal{H}} \left\{ \int H(p) \, d\pi^\mu(p) - H(q) \right\}.
\end{equation}
This class of cost functions plays an important role in \cite{bloedel2021}; a notable example is the channel-capacity cost \citep[e.g.,][]{woodford2012, nimark2019}.
Max-KL cost functions are particular instances of such costs. Observe that dilution linearity holds under any such cost function. Therefore, by the first part of Theorem~\ref{prop:d-linear}, the intersection of Max-Rényi costs and cost functions of the form~\eqref{eq:max-PS} consists only of Max-KL cost functions.

Another approach is to define the cost as a convex monotone transformation of a posterior-separable cost, i.e.,
\begin{equation}\label{eq:convex-PS}
C(\mu) = c\left( \int H(p) \, d\pi^\mu(p) - H(q) \right),
\end{equation}
for some convex and increasing function \( c \). This form was advocated by \cite{denti2022} as a tractable formulation of mixture-convex costs, and has been used in applications \citep[e.g.,][]{zhong2022}.
The Rényi cost function is a particular instance, where
\[
H(p) = 1 - \prod_{i \in \Theta} \left( \frac{p_i}{q_i} \right)^{\alpha_i}, \quad \text{and} \quad c(x) =  \frac{\lambda }{\alpha_{\rm max} - 1} \log(1 - x).
\]
Indeed, this can be seen as a convex transformation of the posterior-separable cost in \cite{Baker}.\footnote{More precisely, \cite{Baker} characterizes cost functions of this form where the parameter \( \alpha \in \mathbb{R}^\Theta \) satisfies \( \sum_i \alpha_i = 1 \), and \( \prod_i x_i^{\alpha_i} - \frac{1}{|\Theta|} \sum_i x_i \) is a convex function of \( x \in (0,1)^\Theta \).}
Observe that any such cost function satisfies independence. Therefore, by the second part of Theorem~\ref{prop:d-linear}, Rényi cost functions can be characterized as the intersection of Max-Rényi costs and cost functions of the form~\eqref{eq:convex-PS}.

\begin{remark}\rm \
Proposition~\ref{prop:additive-linear} implies that KL costs are the only posterior-separable cost functions that satisfy both sub-additivity and identity additivity.
There are other posterior-separable costs that satisfy sub-additivity, including the entropy cost, as we discuss in Appendix~\ref{app:UPS}.
\end{remark}

\section{Application}\label{sec:application}
This section applies Max-Rényi cost functions to information acquisition problems and highlights the role of mixture convexity. We consider DM who holds a prior belief \( q \in \Delta(\Theta) \) over states and has a finite action set \( A \). DM’s payoff from choosing action \( a \in A \) in state \( \theta \in \Theta \) is given by \( u(a,\theta) \). Prior to choosing an action, the DM may acquire any experiment \( \mu \) at cost \( C(\mu) \).\footnote{While our axiomatic analysis excludes unbounded experiments, Max-Rényi cost functions can be uniquely extended to such experiments by approximating them with sequences of bounded experiments.} 
When \( C \) is Blackwell monotone, it suffices to restrict attention to experiments with signal set \( S = A \), such that the DM chooses the action matching the realized signal. The DM’s problem can thus be formulated as maximizing over stochastic choice functions \( \mu: \Theta \to \Delta(A) \):\footnote{An alternative formulation assumes the DM maximizes \( \sum_{\theta \in \Theta} q(\theta) \sum_{a \in A} \mu_\theta(a) u(a,\theta) \) subject to \( C(\mu) \leq K \) for some capacity constraint \( K > 0 \). However, as discussed by \cite{dean2023}, this approach is at odds with empirical evidence suggesting that information acquisition is sensitive to the stakes.}
\[
\max_{\mu: \Theta \to \Delta(A)} \sum_{\theta \in \Theta} q(\theta) \sum_{a \in A} \mu_\theta(a) u(a,\theta) - C(\mu).
\]

Let \({\rm supp}(\mu):=\{a \in A : \sum_{\theta \in \Theta} \mu_\theta(a) > 0\}\) denote the set of actions chosen with positive probability under \(\mu\). Under any posterior-separable cost, Proposition 4 in \cite{denti2022} shows that  DM’s optimal policy requires no more than \(|\Theta|\) actions, i.e., \(\left| {\rm supp}(\mu) \right| \leq |\Theta|\), based on an extreme-point argument from \cite{winkler1988}. This prediction may appear restrictive. When the action set \(A\) is relatively large, it seems reasonable to allow for the possibility that DM invokes more actions than states, i.e., \(\left| {\rm supp}(\mu) \right| > |\Theta|\). 
As noted by \cite{denti2022}, such choice patterns can arise under strictly mixture convex costs. Intuitively, convexity increases the relative value of ``balanced” choice patterns that involve many actions, as opposed to ``extreme” ones concentrated on at most \(|\Theta|\) actions.

To illustrate, consider binary states \(\Theta = \{0,1\}\) with uniform prior \(q(0)=q(1)\). Let the action set be \(A = \{0,1,\phi\}\), and define the utility function as
\[
u(0,0) = u(1,1) = v > w = u(\phi, 0) = u(\phi, 1) > 0 = u(0,1) = u(1,0).
\]
Here, \(v\) is the payoff from matching the action to the state, and \(w\) is the payoff from choosing the safe action \(\phi\), which yields the same payoff across states.

We now apply two special cases of our representations, Max-KL cost and Rényi cost, assumed symmetric for simplicity, as well as posterior-separable costs.\footnote{\cite{bloedel2025} also analyze analogous decision problems using their cost functions based on \(f\)-informativeness.}

\begin{claim}\label{cl:ex}

\

\begin{enumerate}
\item Suppose \(C\) is a symmetric Rényi cost (i.e., \(\alpha_0 = \alpha_1\)). Then there exists \(\overline v\) such that for any \(v \geq \overline v\), there is an interval \(W \subseteq \mathbb{R}\) such that for all \(w \in W\), any optimal \(\mu\) under \((v, w)\) satisfies \({\rm supp}(\mu) = A\).

\item Suppose \(C\) is either a posterior-separable cost or a symmetric Max-KL cost (i.e., \((b, b') \in B \iff (b', b) \in B\)). Then for every \(w > 0\), there exists at most one value \(v > w\) for which there is an optimal policy \(\mu\) with \({\rm supp}(\mu) = A\).
\end{enumerate}

\end{claim}

Under symmetric Rényi costs, DM’s optimal policy features \(\left| {\rm supp}(\mu) \right| = 3\) for a nontrivial set of parameters. In contrast, under posterior-separable costs, the condition \(\left| {\rm supp}(\mu) \right| = 3\) is generically suboptimal. The second part also shows that the predictions under symmetric Max-KL costs closely parallel those of posterior-separable costs. While the last finding is specific to the symmetric setting, it highlights the role of dilution linearity. In particular, the optimal symmetric policies can be written as mixtures of ``symmetric matching” (i.e., \(\mu_0(0) = \mu_1(1) = r\) and \(\mu_0(1) = \mu_1(0) = 1 - r\) for some \(r \in [0,1]\)) and ``fully uninformative choice” (i.e., \(\mu_0(\phi) = \mu_1(\phi) = 1\)). The cost is then linear in the mixture weight due to dilution linearity.

Experiments 1.1 and 2.1 in \cite{dean2023} share the structure of the current decision problem. In these experiments, each subject’s stochastic choice rule \(\mu\) was elicited via repeated decision tasks. The data show that 64\% (resp. 69\%) of subjects in Experiment 1.1 (resp. Experiment 2.1) used all three actions, i.e., \(\left| {\rm supp}(\mu) \right| = 3\), which aligns with the prediction under Rényi costs.\footnote{Data are available at https://dataverse.harvard.edu/dataverse/experimental-tests-of-rational-inattention. \cite{dean2023} used these experiments to demonstrate non-monotone behavior, a general prediction of rational inattention models \citep{matvejka2015}.} 

\section{Discussions}\label{sec:discussion}

\subsection{Pairwise Blackwell Monotonicity}
This section discusses cost functions that are monotone with respect to orders more permissive than the Blackwell order. 
As motivation, observe that the Max-KL cost of an experiment depends only on the information used to distinguish between {\it pairs} of states. This implies that the cost function is not only Blackwell monotone, but also monotone with respect to the Blackwell order restricted to binary state spaces. To formalize this, for any experiment \(\mu = (S, (\mu_i)_{i \in \Theta})\) and any distinct pair \(i, j \in \Theta\), let \(\mu \vert \{i, j\} := (S, (\mu_k)_{k \in \{i, j\}})\) denote the restriction of \(\mu\) to the state space \(\{i, j\}\). We say that \(\mu\) \textit{\textbf{pairwise Blackwell dominates}} \(\nu\) if \(\mu \vert \{i, j\}\) Blackwell dominates \(\nu \vert \{i, j\}\) for every pair \(i, j \in \Theta\). A cost function is said to be \textit{\textbf{pairwise Blackwell monotone}} if \(C(\mu) \geq C(\nu)\) whenever \(\mu\) pairwise Blackwell dominates \(\nu\). When \(|\Theta| > 2\), this axiom is strictly stronger than standard Blackwell monotonicity. 

Pairwise Blackwell dominance is relevant in certain environments. For example, in the case of experiments satisfying the monotone likelihood ratio property, \(\mu\) is more accurate than \(\nu\) in the sense of \cite{lehmann1988} if and only if \(\mu\) pairwise Blackwell dominates \(\nu\) \citep[see][]{jewitt2007}. Moreover, \cite{persico2000} shows that Lehmann’s order characterizes the comparison of information values in certain monotone decision problems.

We now show that under pairwise Blackwell monotonicity, the cost function admits a representation based solely on Rényi divergences between pairs of states. To state this result, we introduce some notation. For each \(i, j \in \Theta\), let \({\mathcal M}_{ij}\) denote the space of Borel measures on \((0,1]\), and define \({\mathcal M} := \prod_{i, j \in \Theta} {\mathcal M}_{ij}\). Thus, each element of \({\mathcal M}\) is a tuple \(m = (m_{ij})_{i, j \in \Theta}\) of measures.

\begin{proposition}\label{thm:pairwise}  
A cost function \(C\) is pairwise Blackwell monotone, mixture convex, sub-additive, and identity additive if and only if there exists a compact set \(M \subseteq {\mathcal M}\) such that for every experiment \(\mu\),
\[
C(\mu) = \max_{m \in M} \sum_{i, j \in \Theta} \int R_t(\mu_i\|\mu_j) \, d m_{ij}(t).
\]  
\end{proposition}

The proposition identifies the class of cost functions that extend Max-KL costs by allowing KL divergence to be replaced with Rényi divergences. According to Theorem~\ref{prop:d-linear}, this broader class corresponds to relaxing the requirement of dilution linearity.

\subsection{Robustness Interpretation of Main Axioms}\label{sec:robust}

Several main axioms used in the paper can be justified by appealing to robustness against manipulations, in the spirit of \cite{bloedel2021}. To illustrate this, suppose that mixture convexity fails, i.e., \(C(\alpha \mu + (1-\alpha) \nu) > \alpha C(\mu) + (1-\alpha) C(\nu)\). Then DM finds it strictly cheaper, in expectation, to randomize between acquiring \(\mu\) and \(\nu\) than to acquire their mixture \(\alpha \mu + (1-\alpha) \nu\). 
Likewise, if sub-additivity fails, i.e., \(C(\mu \otimes \nu) >  C(\mu) + C(\nu)\), DM could reduce costs by purchasing \(\mu\) and \(\nu\) separately, rather than as a bundled experiment \(\mu \otimes \nu\). 

As noted by \cite{pomatto2023}, dilution linearity can also be interpreted in this way when DM can sequentially perform experiments. First, observe that mixture convexity implies \(C(\alpha \mu+(1-\alpha)\phi) \leq \alpha C(\mu)\). Suppose dilution linearity fails, so that \(C(\alpha \mu+(1-\alpha)\phi) < \alpha C(\mu)\). Then, rather than acquiring \(\mu\) directly, DM could repeatedly acquire its diluted version \(\alpha \mu+(1-\alpha)\phi\) until it yields information, reducing the expected total cost to
\[
\sum_{t=0}^\infty (1-\alpha)^t C(\alpha \mu+(1-\alpha)\phi) = \frac{1}{\alpha} C(\alpha \mu+(1-\alpha)\phi).
\]

\cite{bloedel2021} characterize cost functions that are robust to manipulations through all possible sequential experimentation strategies. Such cost functions satisfy both mixture convexity and dilution linearity, consistent with the above reasoning. Moreover, an earlier version of their paper \citep{bloedel2020} considers robustness to more general forms of manipulation, including simultaneous acquisition of distinct experiments, which in turn implies sub-additivity.

\newpage
\appendix

\section{Preliminaries}\label{app:prelim}

This section generalizes the (extended) Rényi divergence, which will be used in Appendix. First, we allow $\alpha$ to take values in $\alpha\in A_+\cup A_-$, 
where \[
A_-  :=  \cup_{k\in\Theta} \{ \alpha \in {\mathbb R}^\Theta : \sum_{i\in\Theta}\alpha_i=1, \alpha_k \geq 1 \}.
\]
Second, we allow divergence parameters to take infinite values. Specifically, we consider divergences of the form
\[
D_\psi^\infty (\mu) :=  \log \; {\rm esssup}_{s \in S} \prod_{i\in\Theta} \frac{d\mu_i}{d\lambda} (s)^{\psi_i}, 
\]
where $\psi$ belongs to 
\[
\Psi := \cup_{k\in\Theta} \{ \psi \in \R^\Theta : \sum_{i\in\Theta} \psi_i = 0,  \psi_k = 1  \}.
\]

To unify these divergences, we employ a different parameterization of divergences, as given by
\[ D_{\gamma,\psi}(\mu)  := \left\{ \begin{array}{cc}
\frac{1}{\max_{i}\alpha^{\gamma,\psi}_i  -1} \log \; \int \prod_{i\in\Theta} \left(\frac{d\mu_i(s)}{d\lambda(s)}\right)^{\alpha^{\gamma,\psi}_i}d \lambda(s)  & \text{ for } \gamma \neq 1, \infty \\
 \sum_{\ell \neq k}-\psi_\ell{\rm KL}(\mu_k \vert \vert \mu_\ell) & \text{ for } \gamma = 1, \; \psi_k=1 \\
D_\psi^\infty (\mu) & \text{ for } \gamma = \infty,
\end{array} \right. \]
where for each $\psi \in \Psi$ with $\psi_k = 1$ and for $\gamma > 0$, we define $\alpha^{\gamma, \psi} := e_k + (\gamma-1) \psi$. 
Note that $\alpha^{\gamma, \psi} \in A_+$ for $\gamma < 1$, $\alpha^{\gamma, \psi} \in A_-$ for $\gamma > 1$. 
Moreover, notice that $\alpha \in A_+ \cup A_-$ if and only if there exists $ \gamma, \psi$ such that $\alpha = \alpha^{\gamma, \psi}$. We can restrict attention to $\gamma \geq \frac{1}{|\Theta|}$. This works since $D_{\alpha^{\gamma , \psi}}$ for $\gamma < \frac{1}{|\Theta|}$ equals $D_{\alpha^{\gamma', \psi'}}$ for some $\psi'$ with different $k$ than $\beta$, and $\gamma ' > \frac{1}{|\Theta|}$. In this manner, we can  collect all divergences as functions $D :  [\frac{1}{|\Theta|}, \infty] \times \Psi \rightarrow \R_+$, whose domain is compact metrizable.



The following lemma collects basic properties of the divergence:

\begin{lemma}\label{lem:approximation} \

\begin{enumerate}
\item  $D$ is additive, i.e., $D_{\gamma,\psi}({\mu\otimes \nu})=D_{\gamma,\psi}({\mu})+D_{\gamma,\psi}({\nu})$, and Blackwell monotone, i.e.,  $D_{\gamma,\psi}({\mu})\geq D_{\gamma,\psi}({\nu})$ whenever $\mu$ Blackwell dominates $\nu$. 
\item For every $\mu\in\cal E$, $D_{\gamma,\psi}(\mu)$ is continuous in $(\gamma,\psi)$.
\item For every $\mu\in\cal E$, there exist sequences $(\overline\mu^k), (\underline\mu^k)$ of finite-signal experiments such that each $\overline\mu^k$ Blackwell-dominates $\mu$, $\mu$ Blackwell-dominates each $\underline\mu^k$,  and $\lim_{k\to\infty}D_{\gamma,\psi}(\underline\mu^k)=\lim_{k\to\infty}D_{\gamma,\psi}(\overline\mu^k)=D(\mu)$ uniformly in $(\gamma,\psi)$.
\end{enumerate}

\end{lemma}

We say $\mu$ is \textit{\textbf{degenerate}} if there exists states $i, j$ such that $\log \frac{d \mu_i}{d \mu_j}$ is 0 $\lambda$-a.s. If $\mu$ is non-degenerate, $D(\mu)$ is bounded away from 0, because $D(\mu) : [\frac{1}{|\Theta|}, \infty] \times \Psi \rightarrow \R$ is continuous and the domain is compact. 

As in Section~\ref{sec:PS}, given an arbitrary full-support prior $q\in\Delta(\Theta)$ one can write divergences as a function of the distribution $\pi\in\Delta (\Delta(\Theta))$ of induced posterior beliefs.  That is,
\[
D_{\gamma,\psi}(\pi)= 
\begin{cases}
\frac{1}{\max_i\alpha^{\gamma,\psi}_i-1} \int \prod_{i\in\Theta}\left(\frac{p_i}{q_i}\right)^{\alpha^{\gamma,\psi}_i}d\pi(p) \text{ for } \gamma\not= 1,\infty  \\
\int \sum_{j\not=i}-\psi_{j}\left(\frac{p_i}{q_i} \log\frac{p_i}{p_j}- \log\frac{q_i}{q_j}\right) d\pi(p) \text{ for } \gamma= 1 \text{ and } \psi_i=1 \\
{\rm ess}\sup_p \log\prod_i \left(\frac{p(i)}{q(i)}\right)^{\psi_i} \text{ for } \gamma= \infty.  \\
 \end{cases}
\]

\subsection{Proof of Lemma~\ref{lem:approximation}}

\noindent{\bf First part:} Additivity follows directly from the functional form of the divergences. 
Blackwell monotonicity can be verified by writing the divergences, as a function of the distributions over posteriors, based on the fact that the distribution induced by $\mu$ is a mean-preserving spread of the one induced by $\nu$.

\noindent{\bf Third part:} We treat each divergence as a function of the distribution over posteriors given fixed full-support prior $q$. Let $\pi$ denote the distribution induced by $\mu$. We show that there exist sequences $(\underline\pi^k)$ and $(\overline\pi^k)$ in $\Delta (\Delta (\Theta))$ with finite support that approximate $\pi$.

We partition $[0,1]$ into $k$ intervals
\[ [0,1]_k := \left\{ \left[0, \frac{1}{k} \right)  ,  \left[\frac{1}{k}, \frac{2}{k}\right) , \ldots , \left[\frac{k-2}{k}, \frac{k-1}{k} \right), \left[\frac{k-1}{k}, 1\right] \right\} \; , \]
and let $[0,1]_k^{|\Theta|}$ be the Cartesian product. Based on this, we partition $\Delta(\Theta)$ by
\[
{\cal B}:=\{ E\cap\Delta(\Theta): E\in [0,1]_k^{|\Theta|}  \}.
\]

For each $B\in\cal B$, let $p^B := \frac{1}{ \pi (B)} \int_{B} p \; \d \pi (p)$ be the expected posterior conditional on $B$ under $\pi$. We define $\underline\pi^k \in \Delta (\Delta (\Theta))$ that is supported on $\{ p^B : B\in \cal B\}$, where 
\[
\underline\pi^k(\{p^B\})=\pi(B)
\]
for each $B\in\cal B$. By construction, $\underline\pi^k$ is a mean-preserving contraction of $\pi^k$, which can be induced by some finite-signal experiment $\underline\mu^k$. 


%

Note that every $B\in \cal B$ is a convex polyhedra, so that it has a finite set of extreme points, ${\rm ext}(B)$. Thus for all $B \in \cal B$ and $p \in B$, we can write $p = \sum_{p' \in {\rm ext}(B)}a_{p'}^B (p) \cdot p'$, where $a_{p'}^B (p) \geq 0$ and $\sum_{p' \in {\rm ext}(B)} a_{p'}^B (p) =1$. Then we define $\overline\pi^k\in \Delta (\Delta (\Theta))$ that is supported on $\cup_{B\in\cal B}{\rm ext}(B)$, where for each $p'\in \cup_{B\in\cal B}{\rm ext}(B)$,
\[ \overline\pi^k (\{p'\}) = \sum_{B\in{\cal B} : p' \in {\rm ext} (B)} \int_B a_{p'}^B (p) \; \d \pi (p) \; . \]
 By construction, $\overline\pi^k$ is a mean-preserving spread of $\pi^k$, which can be induced by some finite-signal experiment $\overline\mu^k$.

We now fix any $\varepsilon\in (0,1)$ and show the existence of $K$ such that $\left\vert D_{\gamma,\psi} (\overline\pi^k) - D_{\gamma,\psi} (\underline\pi^k) \right\vert\leq\frac{\varepsilon}{1-\varepsilon}$ for all $k\geq K$ and $(\gamma,\psi)$. Since $ D_{\gamma,\psi} (\overline\pi^k) \geq  D_{\gamma,\psi} (\pi)\geq  D_{\gamma,\psi} (\underline\pi^k)$ by Blackwell monotonicity in the first pat of the lemma,  this establishes the desired claim. 

Since $\pi$ is induced by a bounded experiment,  there exist $m \in (0,1)$ and $K \in \N$ such that for all $k \geq K$, $p \in [m, 1-m]^{|\Theta|}$ almost surely under $\pi$, $\underline\pi^k$, and $\overline\pi^k$.
Then for all  $p, p' \in [m ,1-m]^{|\Theta|}$ with $\vert \vert p- p' \vert \vert_\infty \leq \delta$, we have that
\begin{eqnarray*}
  \left \vert \log \prod_j p(j)^{\beta_j}  - \log \prod_j  p'(j)^{ \beta_j } \right \vert & \leq & \sum_j \left\vert \beta_j \log \frac{p(j)}{p' (j)} \right\vert \\
  & \leq & \log \frac{m + \delta}{m } + \sum_{j \neq i} \beta_j \log \frac{m }{m + \delta } \\
  & = & \log \frac{(m+ \delta)^2}{m } \rightarrow_{\delta \rightarrow 0} 0,
\end{eqnarray*}
where $i$ is such that $\beta_i=1$.  Importantly, the bound is independent of $\beta$. Therefore there exists $K$ such that for all $k \geq K$ and  $\beta \in \Psi$, 
it holds that 
\[
\sup_{B\in{\cal B}, p'\in {\rm ext}(B)} \prod_i \left( \frac{p^B(i)}{p'(i)}\right)^{\beta_i}  \in [(1+\varepsilon)^{-1}, 1+ \varepsilon].
\]


Take any $k \geq K$. Then for every $(\gamma,\psi)$ with $\gamma<1$,
\begin{eqnarray*}
\left\vert D_{\gamma,\psi} (\overline\pi^k) - D_{\gamma,\psi} (\underline\pi^k) \right\vert & = & \frac{1}{\max_j\alpha^{\gamma,\psi}_j-1}  \log \frac{\int \prod_j  p (j)^{\alpha^{\gamma,\psi}_j} \; \d \overline\pi^k(p)}{\int \prod_j  p (j)^{\alpha^{\gamma,\psi}_j} \; \d \underline\pi^k (p)}  \\
&=&  \frac{1}{\alpha^{\gamma,\psi}_i-1}  \log \frac{\sum_B \int_B \sum_{p'\in {\rm ext}(B)} a_{p'}^B(p) \prod_j  p'(j)^{\alpha^{\gamma,\psi}_j} \; \d \pi(p)}{\sum_B  \prod_j  p^B (j)^{\alpha^{\gamma,\psi}_j} \pi (B)} \\
&\leq&  \frac{1}{\alpha^{\gamma,\psi}_i-1}   \log \sum_B \frac{\int_B \sum_{p'\in {\rm ext}(B)} a_{p'}^B(p) \prod_j  p'(j)^{\alpha^{\gamma,\psi}_j} \; \d \pi(p)}{ \prod_j  p^B (j)^{\alpha^{\gamma,\psi}_j} \pi (B)} \\
&=&  \frac{1}{\alpha^{\gamma,\psi}_i-1}   \log \sum_B \int_B \sum_{p'\in {\rm ext}(B)} a_{p'}^B(p) \prod_j  \left(\frac{p'(j)}{p^B(j)}\right)^{\alpha^{\gamma,\psi}_j} \; \d \pi(p)\\
& = & \frac{1}{\alpha^{\gamma,\psi}_i-1}  \log \sum_{B} \int_B \sum_{p'\in {\rm ext}(B)} a_{p'}^B(p)\left(  \prod_j \left( \frac{p' (j)}{p^B (j)}\right)^{\beta_j} \right)^{\alpha^{\gamma,\psi}_{i}-1}   \frac{p'(i)}{p^B (i)}  \d \pi(p) \\
& \leq & \frac{1}{\alpha^{\gamma,\psi}_i-1}    \log  \sum_{B} \int_B \sum_{p'\in {\rm ext}(B)} a_{p'}^B(p)(1+\varepsilon)^{\alpha^{\gamma,\psi}_{i}-1}\frac{p'(i)}{p^B (i)}  \d \pi(p) \\
& = & \frac{1}{\alpha^{\gamma,\psi}_i-1}   \log (1+\varepsilon)^{\alpha^{\gamma,\psi}_{i}-1} \\
& = & \log (1+\varepsilon) < \varepsilon \;,
\end{eqnarray*}
where $\beta$ and $i$ are such that $\beta_i = 1$, $\alpha_i^{\gamma,\psi}= \max_j \alpha^{\gamma,\psi}_j$, and $\beta_j= \frac{\alpha^{\gamma,\psi}_j}{\alpha^{\gamma,\psi}_{i}-1}$ for $j \neq i$.

Similarly,  for every $(\gamma,\psi)$ with $\gamma>1$,
\begin{eqnarray*}
\left\vert D_{\gamma,\psi} (\overline\pi^k) - D_{\gamma,\psi} (\underline\pi^k) \right\vert & = & \frac{1}{1-\alpha^{\gamma,\psi}_{i} }  \log \frac{\int \prod_j  p (j)^{\alpha^{\gamma,\psi}_j} \; \d \underline\pi^k(p)}{\int \prod_j  p (j)^{\alpha^{\gamma,\psi}_j} \; \d \overline\pi^k (p)}  \\
& \leq  & \frac{1}{1-\alpha^{\gamma,\psi}_{i} }  \log \sum_B \frac{ \prod_j  p^B (j)^{\alpha^{\gamma,\psi}_j} \pi (B)} {\int_B \sum_{p'\in {\rm ext}(B)} a_{p'}^B(p) \prod_j  p'(j)^{\alpha^{\gamma,\psi}_j} \; \d \pi(p)} \\
& = & \frac{1}{1-\alpha^{\gamma,\psi}_{i} }  \log \sum_{B} \left(  \int_B  \sum_{p'\in {\rm ext}(B)} a_{p'}^B(p)\prod_j  \left( \frac{p (j)}{p^B (j)}\right)^{\alpha^{\gamma,\psi}_j} \d \pi(p) \right)^{-1} \\
& = &\frac{1}{1-\alpha^{\gamma,\psi}_{i} }   \log \sum_{B} \left( \int_B  \sum_{p'\in {\rm ext}(B)} a_{p'}^B(p)\left( \prod_j \left( \frac{p^B (j)}{p' (j)}\right)^{\beta_j} \right)^{1-\alpha^{\gamma,\psi}_{i}}  \frac{p'(i)}{p^B (i)} \d \pi(p) \right)^{-1} \\
& \leq & \frac{1}{1-\alpha^{\gamma,\psi}_{i} }  \log  \sum_{B}  \left( \int_B  \sum_{p'\in {\rm ext}(B)} a_{p'}^B(p)(1+\varepsilon)^{\alpha^{\gamma,\psi}_{i}-1}  \frac{p'(i)}{p^B (i)} \d \pi(p) \right)^{-1} \\
& = & \frac{1}{1-\alpha^{\gamma,\psi}_{i} }  \log (1+\varepsilon)^{1-\alpha^{\gamma,\psi}_{i}} \\
& = & \log (1+\varepsilon)<\varepsilon,
\end{eqnarray*}
where $\beta$ and $i$ are defined as before.

Since $\underline\pi^k$ and $\overline\pi^k$ are induced by finite-signal experiments, Proposition 17 of \cite{farooq2024} shows that $D_{\gamma,\psi}(\underline\pi^k)$ and $D_{\gamma,\psi}(\underline\pi^k)$ are continuous in $\gamma$.  Thus the above observations ensure that $\left\vert D_{\gamma,\psi} (\overline\pi^k) - D_{\gamma,\psi} (\underline\pi^k) \right\vert\leq \varepsilon$ at every $(\gamma,\psi)$ with $\gamma=1$.

For any $\psi \in \Psi$ and any $p'$ in the support of $\overline\pi^k$, there is some $p^B$ in the support of $\underline\pi^k$ such that  $\frac{ \prod_i p^B(i)^{\psi_i}}{ \prod_i p'(\theta)^{\psi_i}}  \geq (1+\varepsilon)^{-1}$, which implies $D^\infty_\psi(\overline\pi^k)\leq D^\infty_\psi(\underline\pi^k)+\log (1+\varepsilon)$. Combined with a symmetric argument, we have  $\vert D_\psi^\infty (\overline\pi^k) - D_\psi^\infty (\underline\pi^k) \vert \leq \log (1+ \varepsilon)<\varepsilon$.

\

\noindent{\bf Second part:} Observe first that $D_{\gamma,\psi}(\mu)$ is continuous in $\psi$ at every $\mu$. 
Proposition 17 of \cite{farooq2024} shows that $D_{\gamma,\psi}(\mu)$ is continuous in $\gamma$ at every finite-signal experiment $\mu$.
For general $\mu$, we take sequences $(\underline\mu^k), (\overline\mu^k)$ of finite-signal experiments from the third part of the lemma. Then, for all $(\gamma,\psi)$ and $k$, 
\[ D_{{\gamma, \psi}} (\underline\mu^k) \leq D_{{\gamma, \psi}} (\mu) \leq D_{{\gamma, \psi}} (\overline\mu^k) \; , \]
which thus implies that for all $k$ and $\gamma^*$, 
\[ D_{\gamma^*,\psi} (\underline\mu^k) \leq \liminf_{\gamma \rightarrow \gamma^*} D_{{\gamma, \psi}} (\mu)\leq \limsup_{\gamma \rightarrow \gamma^*} D_{{\gamma, \psi}} (\mu)\leq D_{\gamma^*,\psi}(\overline\mu^k) \]
Thus by taking the limit $k\to\infty$, we obtain $\lim_{\gamma \rightarrow \gamma^*} D_{{\gamma, \psi}} (\mu)=D_{{\gamma^*, \psi}} (\mu)$, as desired. 

%

\section{Proofs of Main Results}

\subsection{Generalization of Theorem~\ref{thm:subadditive}}

To prove Theorem~\ref{thm:subadditive}, we first characterize a representation based on the generalized divergence $D_{\gamma,\psi}$ introduced in Appendix~\ref{app:prelim}.  By abusing notation, let $\calM$ be the set of Borel measures on $[\frac{1}{|\Theta|}, \infty] \times \Psi$.

\begin{theorem}\label{thm:general}
$C$ is Blackwell monotone, sub-additive, and identity additive if and only if there exists compact set $M \subseteq \calM$ such that
\begin{equation}\label{eq:general}
C (\mu) = \max_{m \in M} \int D_{\gamma , \psi}(\mu) \; \d m (\gamma, \psi).
\end{equation}
\end{theorem}

This representation is more general than Max-Rény cost, as it need not satisfy mixture convexity.  This is reduced to Max-Rény cost if every measure $m\in M$ is supported on $[\frac{1}{|\Theta|}, 1]\times\Psi$.

\subsubsection{Proof of Theorem~\ref{thm:general}}

``if'' direction is clear from the functional form of $C$, using additivity and Blackwell monotonicity of generalized Rényi divergence. Below we prove ``only if'' direction.  We fix a cost function $C$ that is Blackwell monotone, sub-additive, and identity  additive.

We start by restricting attention to finite experiments $\calE_f \subset \calE$.
We first show that the cost function is monotone in divergences:
\begin{lemma}\label{lem:monotone}
For any $\mu,\nu\in \calE_f$ such that $D(\mu)\geq D(\nu)$, we have $C(\mu)\geq C(\nu)$. 
\end{lemma}
\begin{proof}
Suppose first that $D(\mu) > D(\nu)$. By Theorem 19 in \cite{farooq2024}, there exists $k$ such that $\mu^{\otimes k}$ Blackwell dominates $\nu^{\otimes k}$, and thus, by identity additivity  and Blackwell monotonicity,
\[ C (\mu) = \frac{1}{k} C (\mu^{\otimes k}) \geq \frac{1}{k} C (\nu^{\otimes k}) = C (\nu). \]
Suppose now that $D(\mu)\geq D(\nu)$. 
Take any non-degenerate $\phi\in{\cal E}_f$, so that $D(\phi) > 0$. Then for every $k >0$, $\gamma \in [\frac{1}{|\Theta|}, \infty]$ and $\psi \in \Psi$,
\[ D(\mu^{\otimes k} \otimes \phi) (\gamma, \beta) = k D_{\gamma,\psi}(\mu)  + D_{\gamma,\psi}( \phi) > k D_{\gamma,\psi}(\nu) = D_{\gamma,\psi}(\nu^{\otimes k}), \]
so that $C (\mu^{\otimes k} \otimes \phi) \geq C (\nu^{\otimes k})$. Thus, by identity additivity and sub-additivity,
\[  C (\mu) + \frac{1}{k} C (\phi)=\frac{1}{k} (C(\mu^{\otimes k})+C(\phi)) \geq \frac{1}{k} (C(\mu^{\otimes k}\otimes \phi)) \geq \frac{1}{k} (C(\nu^{\otimes k}))= C (\nu) . \]
As it holds for every $k$, we have $C(\mu)\geq C(\nu)$.
\end{proof}

Let $X$ denote the space of all continuous functions $x:[\frac{1}{|\Theta|}, \infty] \times \Psi \to \mathbb R$, endowed with the sup norm $\|\cdot\|_\infty$. It's norm dual is identified with the space of finite Borel measures on $[\frac{1}{|\Theta|}, \infty] \times \Psi$ (e.g., Corollary 14.15 in \cite{aliprantis}). Let
\[ \mathcal{H} := \left\{ x \in X : \exists \mu \in \mathcal{E}_f, \; x = D(\mu) \right\} \]
and 
\[ \mathrm{cone} (\mathcal{H}) := \left\{ \left. \sum_{i=1}^n \alpha_i D({\mu^i}) \right\vert n\in{\mathbb N}, \; \forall i=1,\ldots,n, \; \alpha_i > 0, \; \mu^i \in \mathcal{E}_f \right\}. \]

The lemma below shows that the cost function can be represented by a functional on $\mathrm{cone} (\mathcal{H})$:
\begin{lemma}\label{lem:sub-additive}
There exists a Lipschitz continuous map $F:\mathrm{cone} (\mathcal{H}) \to\mathbb R$ such that 
$C(\mu)=F(D(\mu))$ for each $\mu\in{\cal E}_f$, where $F$ is (i) monotone: $F(x)\geq F(x')$ for $x\geq x'$, (ii) positively homogeneous: $F(\alpha x)=\alpha F(\alpha x)$ for all $x$ and $\alpha>0$, (iii) sub-additive: $F(x+x')\leq F(x)+F(x')$ for all $x,x'$.

\end{lemma}
\begin{proof}
If $D(\mu) = D(\nu)$ then  $C(\mu) = C(\nu)$ by Lemma~\ref{lem:monotone}. Thus there exists $G : \mathcal{H} \rightarrow \mathbb{R}$ such that $C (\mu) = G (D(\mu))$ for each $\mu\in{\cal E}_f$.  $G$ is monotone by Lemma~\ref{lem:monotone}.
$G$ is also sub-additive:
\[ G (D(\mu) + D(\nu)) = G (D({\mu \otimes \nu})) = C (\mu \otimes \nu ) \leq C (\mu) + C (\nu) = G (D(\mu)) + G (D(\nu)) . \]
by additivity of divergences and sub-additivity of $C$.

We show that $G$ is positively homogeneous in $\alpha \in \mathbb{Q}_{+}$: This holds whenever $\alpha\in\mathbb N$, by $D({\mu^{\otimes \alpha}})=\alpha D({\mu})$ and identity additivity. Now consider the general case where $\alpha=\frac{r}{m}$ for some $r,m\in\mathbb N$ and $D(\nu)=\alpha D(\mu)$. By $m D(\nu) = r D(\mu)$ and positive homogeneity of $G$ in integers, we have $m G (D(\nu)) = G (m D(\nu)) = G (r D(\mu)) = r G (D(\mu))$,  and thus
$G(\alpha D(\mu))=G (D(\nu)) = \frac{r}{m} G (D(\mu)) = \alpha G (D(\mu))$.

We construct an extension $F$ of $G$ to
\[ 
\mathrm{cone}_{\mathbb{Q}} (\mathcal{H}) := \left\{ \left. \sum_{i=1}^n \alpha_i D({\mu^i}) \right\vert n\in\mathbb N, \; \forall i=1,\ldots, n, \; \alpha_i \in \mathbb{Q}_{+}, \: \mu^i \in \mathcal{E}_f \right\}=\bigcup_{n \in\mathbb N} \frac{1}{n} \mathcal{H}
 \]
 where the second equality follows by the additivity of divergences. Specifically, we set  $F (\frac{1}{r} D(\mu)) := \frac{1}{r} G (D(\mu))$ for each $\mu\in{\cal E}_f$ and $r\in\mathbb N$. This is well defined since, if there exist $\mu, \nu \in \mathcal{E}_f$ such that $\frac{1}{n} D(\mu) = \frac{1}{m} D(\nu)$, then positive homogeneity of $G$ ensures $F(\frac{1}{n} D(\mu)) = \frac{1}{n} G (D(\mu)) = \frac{1}{m} G (D(\nu)) = F(\frac{1}{m} D(\nu))$. 
Clearly, $F$ inherits monotonicity, sub-additivity, and  positive homogeneity in $\alpha \in \mathbb{Q}_{+}$.

We show that $F$ is Lipschitz continuous: Take any $\mu \in \mathcal{E}_f$ that is non-degenerate. Since $D(\mu)$ is bounded away from 0, there exists $n \geq 1$ such that $\phi = \mu^{\otimes n}$ satisfies $D_{\gamma,\psi}(\phi)=n D_{\gamma,\psi}(\mu) \geq 1$ for every $\gamma$ and $\psi$. For every $x, y \in  \mathrm{cone}_{\mathbb{Q}} (\mathcal{H})$, $\gamma$, and $\psi$, 
\[ \vert x (\gamma, \beta) - y(\gamma, \beta) \vert \leq \| x - y \|_\infty \leq \| x - y \|_\infty D_{\gamma,\psi} (\phi).
\]
Then for every rational number $r \geq \| x - y \|_\infty$,
\[ F (x) \leq F(y + r D(\phi)) \leq F(y) + F(r D(\phi)) = F(y) + r F(D(\phi)) \]
by monotonicity, sub-additivity, and positive homogeneity of $F$.
Analogously, $F (y) \leq F (x) + r F(D(\phi))$,
and thus $ \vert F(x) - F (y) \vert \leq r F (D(\phi)).$
By letting $r \searrow \| x - y \|_\infty$, we obtain $\vert F(x) - F (y) \vert \leq  \| x - y \|_\infty F (D(\phi)).$

Since $F$ is Lipschitz continuous, it can be extended to the closure $\overline{\mathrm{cone}_{\mathbb{Q}} (\mathcal{H})}$,
which contains $\mathrm{cone} (\mathcal{H})$.
We now verify the three properties of  $F$  on $\mathrm{cone} (\mathcal{H})$. To do so, we take any $x, y \in \mathrm{cone} (\mathcal{H})$ and sequences  $(x_n)$, $(y_n)$ in $\mathrm{cone}_{\mathbb{Q}} (\mathcal{H})$ such that $ x_n \rightarrow x$ and $ y_n \rightarrow y$ in sup norm. 

\begin{enumerate}
\item {\it Monotonicity}: Suppose  $x \geq y$. Take any $\mu$ that is non-degenerate. For every $m\in\mathbb Q_{++}$ there exists $N$ such that
\[  x_n - x \geq - \frac{1}{2m} D(\mu), \; y_n - y \leq \frac{1}{2m} D(\mu)  \]
for every $n \geq N$, which ensures $y_n \leq  x_n + \frac{1}{m} D(\mu).$ By monotonicity, sub-additivity, and positive homogeneity on $\mathrm{cone}_{\mathbb{Q}} (\mathcal{H})$,
\[  F ( y_n ) \leq F \left( x_n + \frac{1}{m} D(\mu)\right) \leq F(  x_n) + \frac{1}{m} F( D(\mu)) \]
and thus, taking $n\to\infty$, $F (y) \leq F (x) + \frac{1}{m} F( D(\mu))$.
Since this holds for every $m$, $F(y)\leq F(x)$. 

\item {\it Sub-additivity}: Observe that $x_n + y_n \in \mathrm{cone}_{\mathbb{Q}} (\mathcal{H})$ for every $n$ and $\lim_{n \to\infty}x_n + y_n = x +y$. Thus $F (x+y) = \lim_{n\to\infty} F(x_n + y_n) \leq \lim_{n\to\infty} F (x_n) + F (y_n) = F(x) + F (y)$ by sub-additivity on $\mathrm{cone}_{\mathbb{Q}} (\mathcal{H})$. 

\item {\it Positive homogeneity}: If $\alpha \in \mathbb{Q}_+$, then $\alpha x_n \rightarrow_n \alpha x$, and $\alpha x_n \in \mathrm{cone}_{\mathbb{Q}} (\mathcal{H})$ for every $n$. Then by positive homogeneity (in rationals) on $\mathrm{cone}_{\mathbb{Q}} (\mathcal{H})$,
$ F (\alpha x) = \lim_{n\to\infty} F (\alpha x_n) = \lim_{n\to\infty} \alpha F (x_n) = \alpha F (x)$.
If $\alpha > 0$ is not rational, the same conclusion holds by  taking a sequence of positive rationals $\alpha_n \rightarrow \alpha$ and by Lipschitz continuity of $F$. 
\end{enumerate}
\end{proof}

\begin{lemma}\label{lemma:infinitesignalextension}
The same map $F: \mathrm{cone} (\mathcal{H}) \rightarrow \R$ represents the cost of all experiments with bounded likelihoods, i.e., $C(\mu) = F(D(\mu))$ for all $\mu \in \cal E$.
\end{lemma}

\begin{proof}
As shown in Lemma~\ref{lem:approximation}, given any $\mu \in \cal E$ its divergences can be approximated by sequences of finite-signal experiments with bounded likelihoods $\mu_\pi^k, \mu_\lambda^k \in {\cal E}_f$ such that, for every $k$, $\mu_\lambda^k$ Blackwell dominates $\mu$ which Blackwell dominates $\mu_\pi^k$.
Thus, for every $k$,
\[ F(D(\mu_\pi^k)) = C(\mu_\pi^k) \leq C(\mu) \leq C(\mu_\lambda^k) = F(D(\mu_\lambda^k)) \]
by Blackwell monotonicity of $C$.

Since $D(\mu_\pi^k), D(\mu_\lambda^k) \rightarrow D(\mu)$ uniformly, continuity of $F$ ensures that
\[ F(D(\mu)) = \lim_k F(D(\mu_\pi^k)) \leq C(\mu) \leq \lim_k F(D(\mu_\lambda^k)) = F(D(\mu)) \; , \]
as desired.
\end{proof}

We extend $G$ from $\mathrm{cone} (\mathcal{H})$ to $X$ by 
\[
\tilde G(x):=\inf_{y\in \mathrm{cone} (\mathcal{H}), y\geq x} G(y) \text{ for each } x\in X.
\]
Observe that $\tilde G$ is finite valued; since any $x\in X$ is bounded, there is some $y\in \mathrm{cone} (\mathcal{H})$ that dominates $x$. Moreoever, $\tilde G(x)=G(x)$ for $x\in \mathrm{cone} (\mathcal{H})$, and it inherits  monotonicity, positive homogeneity, subadditivity, lower-semicontinuity of $\tilde G$. 

By Theorem 7.51 of \cite{aliprantis}, there is a closed and convex set $M$  of Borel measures on $[ \frac{1}{|\Theta|}, \infty] \times \Psi$ such that 
\[ 
G (x) = \sup_{m \in M} \int x(\gamma, \psi) d m (\gamma, \psi)
 \]
for each $x\in X$.

We observe $m$ is non-negative for every $m\in M$. If $m(A)<0$ for some Borel $A\subseteq[ \frac{1}{|\Theta|}, \infty] \times \Psi$, then take a sequence $x^n\nearrow -1_{A}$ in $X$, where $1_A$ denotes the indicator function on $A$.  Then $\liminf_{n\to\infty} \tilde G(x^n)\geq \liminf_{n\to\infty} \int x^n(\gamma, \psi) dm(\gamma, \psi)=-m(A)$.  But $\tilde G(x^n)\leq \tilde G(1_\emptyset)=0$ by monotonicity of $\tilde G$, a contradiction. 

Moreover, $\sup_{m\in M}m([ \frac{1}{n}, \infty] \times \Psi)$ is finite; otherwise $\sup_{m\in M}\int x(\gamma, \psi) dm(\gamma, \psi)=\infty$ for any strictly positive $x$. Thus $\sup_{m\in M, \|x\|_\infty=1}\int x (\gamma, \psi)dm(\gamma, \psi)$ is finite, which ensures that $M$ is compact (by Banach-Alaoglu theorem). This allows us to replace ``sup'' with ``max'' in the above representation, so that for every $\mu \in \mathcal{E}$,
\[ C(\mu) = G (D(\mu)) =\max_{m \in M} \int D_{\gamma, \psi}(\mu)  d m (\gamma, \psi) . \]


\hfill$\square$
\subsection{Proof of Theorem~\ref{thm:subadditive}}

We say $C$ is \textit{\textbf{dilution convex}}  if  \(C(\alpha \mu+ (1-\alpha)\phi) \leq \alpha C(\mu) +(1-\alpha)C(\phi)\) for all \(\mu, \phi\) and \(\alpha \in (0,1)\) whenever $\phi$ is uninformative. First, we show that dilution convexity (and thus mixture convexity) of Max-R\'{e}nyi cost $C$ implies that each $m(\cdot, \Psi)$ is supported on $[\frac{1}{|\Theta|}, 1]$.
Take any $\mu$ non-degenerate, and let $\nu^k:= \frac{1}{k} \mu^{\otimes k}$ for every $k \in \mathbb{N}$. For $\gamma \neq 1, \infty$,
\begin{equation}\label{eq:product}
D_{\gamma,\psi} (\nu^k) = \frac{1}{\gamma-1} \log \left( \frac{k-1}{k} + \frac{1}{k} \left(   \int_S \prod_i \left(  \frac{d \mu_i}{d \lambda} (s) \right)^{\alpha_i^{\gamma,\psi}}  d \lambda (s)   \right)^k \right) . \end{equation}
For $\gamma > 1$, $ \int_S \prod_i \left(  \frac{d \mu_i}{d \lambda} (s) \right)^{\alpha_i^{\gamma,\psi}}  d \lambda (s)  > 1$, so that $\lim_{k \rightarrow \infty} D_{\gamma,\psi} (\nu^k) = \infty $. Similarly, $ \lim_{k \rightarrow \infty} D_{\infty,\Psi} (\nu^k) = \lim_{k \rightarrow \infty} k D_{\infty,\psi} (\mu) = \infty$.
Assume by contradiction that $m ((1, \infty]\times\Psi) > 0$ for some $m\in M$. Then $\lim_{k \rightarrow \infty} C(\nu^k) = \infty$. However, by dilution convexity and identity additivity
\[ C(\nu^k) = C \left(\frac{1}{k} \mu^{\otimes k}\right) \leq \frac{1}{k}C \left(\mu^{\otimes k}\right)= C(\mu) < \infty , \]
 a contradiction.

Second, we show that the concentration of measures on $[\frac{1}{|\Theta|},1]$ implies mixture convexity of the Max-R\'{e}nyi cost $C$. Fix experiments $\mu, \nu \in \cal E$ and $\alpha \in (0,1)$. 
As noted in the main text, $D_{\gamma,\psi} (\alpha\mu+(1-\alpha)\nu) \leq \alpha D_{\gamma,\psi} (\mu) + (1-\alpha) D_{\gamma,\psi} (\nu)$.
Then, it follows that
\begin{eqnarray*}
C (\alpha \mu + (1-\alpha) \nu) & = & \max_{m \in M} \int D_{\gamma,\psi} (\alpha\mu+(1-\alpha)\nu) dm(\gamma,\psi) \\
 &  \leq &  \max_{m \in M}  \int \alpha D_{\gamma,\psi} (\mu) + (1-\alpha) D_{\gamma,\psi} (\nu)  dm(\gamma,\psi) \\
 & \leq & \alpha \max_{m \in M} \int D_{\gamma,\psi} (\mu) dm(\gamma,\psi) + (1-\alpha) \max_{m \in M}\int D_{\gamma,\psi} (\nu) dm(\gamma,\psi) \\
 & = & \alpha C(\mu) + (1-\alpha) C (\nu). 
\end{eqnarray*}
\hfill$\square$

\subsection{Proof of Proposition~\ref{prop:additive-linear}}
\subsubsection{First part}
``If'' direction is clear from the functional form of $C$, using additivity and Blackwell monotonicity of the extended Rényi divergence. For ``only if'' direction, take a representation of $C$ based on set of measures $M$ from Theorem~\ref{thm:subadditive}. For each experiment $\mu$, let $M_\mu:=\argmax \int D_{\alpha,\beta}(\mu) dm(\alpha,\beta)$, which is a closed set.  For any finite collection of experiments $\mu^1,\ldots, \mu^n$, their product $\otimes_{\ell=1}^n \mu^\ell$ is defined in an obvious manner. Then the additivity ensures
\[
\max_{m\in M} \int D_{\alpha,\beta}(\otimes_{\ell=1}^n \mu^\ell) dm(\alpha,\beta)=C(\otimes_{\ell=1}^n \mu^\ell)=\sum_{\ell=1}^n C(\mu^\ell)=\sum_{\ell=1}^n\max_{m\in M} \int D_{\alpha,\beta}(\mu^\ell) dm(\alpha,\beta).
\]
This implies $m\in \cap_{\ell=1}^n M_{\mu^\ell}$ for some $m\in M_{\otimes_{\ell=1}^n \mu^\ell}$, as otherwise $C(\otimes_{\ell=1}^n \mu^\ell)<\sum_{\ell=1}^n C(\mu^\ell)$. 
Then, since $M$ is compact, the finite-intersection property ensures $m\in\cap_{\mu}M_\mu$ for some $m$. This guarantees that $C(\mu)=\int D_{\alpha,\beta}(\mu)dm(\alpha,\beta)$ for every $\mu$, so that $C$ admits the representation with $\{m\}$.


\subsubsection{Second Part}
``If'' direction follows from mixture linearity of KL divergence.
 For ``only if'' direction, take a representation of $C$ based on set of measures $M$ from the first part of Theorem~\ref{prop:d-linear}, where $m(\cdot, \Psi)$ is supported on 1 for every $m\in M$. For any finite collection of experiments $\mu^1,\ldots, \mu^n$, their mixture $\sum_{\ell=1}^np_\ell \mu^\ell$ with respect to convex weights $(p_1,\ldots, p_n)$ is defined in an obvious manner. Then the mixture linearity  ensures
\[
\max_{m\in M} \int D_{\alpha,\beta}(\sum_{\ell=1}^np_\ell \mu^\ell) dm(\alpha,\beta)=C(\sum_{\ell=1}^np_\ell \mu^\ell)=\sum_{\ell=1}^n p_\ell C(\mu^\ell)=\sum_{\ell=1}^np_\ell\max_{m\in M} \int D_{\alpha,\beta}(\mu^\ell) dm(\alpha,\beta).
\]
This implies $m\in \cap_{\ell=1}^n M_{\mu^\ell}$ for some $m\in M_{\sum_{\ell=1}^np_\ell \mu^\ell}$, as otherwise $C(\sum_{\ell=1}^np_\ell \mu^\ell)<\sum_{\ell=1}^np_\ell C(\mu^\ell)$. 
Then, since $M$ is compact, the finite-intersection property ensures $m\in\cap_{\mu}M_\mu$ for some $m$. This guarantees that $C(\mu)=\int D_{\alpha,\beta}(\mu)dm(\alpha,\beta)$ for every $\mu$, so that $C$ admits the representation with $\{m\}$, where  $m(\cdot, \Psi)$ is supported on 1. 
\hfill$\square$


\subsubsection{Alternative Characterization of KL Cost}\label{app:KL2}

We show that $C$ is represented by a KL cost if and only if it is Blackwell monotone, dilution linear, and additive.  This is a non-binary state extension of Theorem 5 in \cite{pomatto2023}. 
``Only if'' direction follows from Proposition~\ref{prop:additive-linear}. To show ``if'' direction, first take the general representation from Theorem~\ref{thm:general}.  Since dilution linearity implies dilution convexity, the proof of Theorem~\ref{thm:subadditive} shows that $m(\cdot, \Psi)$ is supported on $[\frac{1}{|\Theta|}, 1]$ for each $m\in M$. Moreover, the proof of the first part of Proposition~\ref{prop:additive-linear} implies $M=\{m\}$.  
As in the proof of Theorem~\ref{thm:subadditive}, take  $\nu^k:= \frac{1}{k} \mu^{\otimes k}$, where $\mu$ is non-degenerate. 
 For $\gamma < 1$, $ \int_S \prod_i \left(  \frac{d \mu_i}{d \lambda} (s) \right)^{\alpha_i^{\gamma,\psi}}  d \lambda (s)  < 1$, so that $\lim_{k \rightarrow \infty} D_{\gamma,\psi} (\nu^k) = 0$ by  (\ref{eq:product}). At the same time, $D_{\gamma,\psi} (\nu^k) $ is independent of $k$ for $\gamma=1$. Thus if $m([\frac{1}{|\Theta|},1)\times\Psi)>0$ then $\lim_{k\to\infty}C(\nu^k)<C(\nu^1)$, which then violates the combination of dilution linearity and additivity. Therefore $m$ is supported on $\{1\}\times\Psi$, as desired.

\subsection{Proof of Theorem~\ref{prop:d-linear}}

\subsubsection{First Part}
 ``If'' direction follows from linearity of KL divergence and positive homogeneity of the maximum.

For ``only if'' direction, take any Max-R\'{e}nyi cost given by the set of measures $M$.
Suppose that there exists $\mu \in \cal E$ and $m \in M$ such that $C(\mu) =  \int D_{\gamma,\psi} (\mu) d m (\gamma,\psi)>0$ and $m ([\frac{1}{|\Theta|},1)\times\Psi) > 0$. Let $\nu^2 := \frac{1}{2} \mu^{\otimes 2}$. Then for each $\gamma<1$ and $\psi$, 
\begin{equation}\label{eq:ineq}
D_{\gamma,\psi} (\nu^2) < D_{\gamma,\psi} (\mu).
\end{equation}
It follows that
\[  \int D_{\gamma,\psi} ( \nu^2) d m (\gamma,\psi) < \int D_{\gamma,\psi} ( \mu) d  m(\gamma,\psi) = C(\mu) . \]
Dilation linearity then requires the existence of $m^2 \in M$, different from $m$, such that
\[  \int D_{\gamma,\psi} ( \nu^2) d m^2 (\gamma,\psi) = C(\mu) . \]
If  $m^2 ([\frac{1}{|\Theta|}, 1)) > 0$, then (\ref{eq:ineq}) would imply the contradiction
\[\int D_{\gamma,\psi} ( \mu) d m^2 (t)  >  \int D_{\gamma,\psi} ( \nu^2) d m^2 (\gamma,\psi) = C(\mu) . \]
Hence,  it must be that $m^2$ is supported on $\{1\}\times\Psi$. Then,
\[ \int D_{\gamma,\psi} ( \mu) d m^2 (t) = C(\mu) , \]
and therefore we can equivalently restrict attention to the (nonempty, convex, and closed) subset of $M$ of measures that are supported on $\{1\}\times\Psi$.
\hfill$\square$

\subsubsection{Second Part}

``If'' direction follows from the observation that $C$ is a monotone transformation of a posterior-separable cost. To prove ``only if'' direction, take $C$ that satisfies independence. 

As discussed in the main text, we can pick any full-support prior $q$ and write $C(\cdot)$, as well as each divergence $D_{\gamma,\psi}(\cdot)$,  as a function of the distribution $\pi$ over posteriors induced from prior $q$. 
For $\gamma<1$, we can write $D_{\gamma,\psi}=\frac{1}{\alpha_{\rm max}^{\gamma,\psi}-1}\log V_{\alpha^{\gamma,\psi}} (\pi)$ , where
\[
V_\alpha(\pi):=\int\prod_{i\in\Theta} \left(\frac{p_i}{q_i}\right)^{\alpha_i} d\pi(p)
\] is linear in $\pi$. $D_{\gamma,\psi}(\pi)$ is linear in $\pi$ when $\gamma=1$. For each $\pi$, let $M_\pi:=\argmax_{m\in M}\int D_{\gamma,\psi}(\pi)dm(\gamma,\psi)$.

Fix any $\pi',\pi''$ such that $C(\pi')=C(\pi'')$. For any $\pi$ and $\varepsilon, \kappa \in (0,1)$, observe that $C((1-\varepsilon)\pi+\varepsilon\kappa \pi'+\varepsilon (1-\kappa)\pi'')$ is independent of $\kappa$ by the independence axiom. Thus 

\begin{align}
0=&\frac{\partial C((1-\varepsilon)\pi+\varepsilon\kappa \pi'+\varepsilon (1-\kappa)\pi'')}{\partial \kappa}\nonumber \\
= &\frac{\partial \int V_{\gamma,\psi}((1-\varepsilon)\pi+\varepsilon\kappa \pi'+\varepsilon (1-\kappa)\pi'')dm_\varepsilon(\gamma,\psi)}{\partial \kappa} \nonumber \\
=&\int_{\gamma<1} \frac{\varepsilon}{\alpha_{\rm max}^{\gamma,\psi}-1} \frac{V_{\gamma,\psi}(\pi')- V_{\gamma,\psi}(\pi'')}{V_{\gamma, \psi}((1-\varepsilon)\pi+\varepsilon\kappa \pi'+\varepsilon (1-\kappa)\pi'')} dm_\varepsilon(\gamma, \psi) \nonumber
 \\
&+ \int_{\gamma=1} \varepsilon \left(D_{\gamma,\psi}(\pi')- D_{\gamma,\psi}(\pi'')\right) dm_\varepsilon(\gamma, \psi) \label{eq:M1}
\end{align}
for some $m_\varepsilon\in M_{(1-\varepsilon)\pi+\varepsilon\kappa \pi'+\varepsilon (1-\kappa)\pi''}$  by the envelope theorem. Moreover,
\begin{align}
0&\geq  \frac{\partial^2 \int((1-\varepsilon)\pi+\varepsilon\kappa \pi'+\varepsilon (1-\kappa)\pi'')dm_\varepsilon(\gamma,\psi)}{(\partial \kappa)^2} \nonumber \\
&=\int_{\gamma<1} \frac{-\varepsilon^2}{\alpha_{\rm max}^{\gamma,\psi}-1} \left(\frac{V_{\gamma,\psi}(\pi')- V_{\gamma,\psi}(\pi'')}{V_{\gamma, \psi}((1-\varepsilon)\pi+\varepsilon\kappa \pi'+\varepsilon (1-\kappa)\pi'')} \right)^2dm_\pi(\gamma, \psi). \label{eq:M2}
\end{align}

To see the inequality, observe that the reversed inequality would imply  
\[
\int((1-\varepsilon)\pi+\varepsilon\kappa' \pi'+\varepsilon (1-\kappa')\pi'')dm_\varepsilon(\gamma,\psi)>C((1-\varepsilon)\pi+\varepsilon\kappa \pi'+\varepsilon (1-\kappa)\pi'')
\]
for $\kappa'$ sufficiently close to $\kappa$, which contradicts the representation.

Since $M$ is compact, the sequence $(m_\varepsilon)$ as $\varepsilon\to 0$ admits a limit $m_\pi$ of its convergent subsequence.  Observe that $m_\pi\in M_\pi$, since the integral $\int D_{\gamma,\psi}(\cdot)dm(\gamma,\psi)$ is continuous in $m$. By dividing (\ref{eq:M1}) by $\varepsilon$ and taking the limit $\varepsilon\to 0$, via the bounded-convergence theorem we obtain 
\begin{equation}\label{eq:ind1'}
0=\int_{\gamma<1} \frac{1}{\alpha_{\rm max}^{\gamma,\psi}-1} \frac{V_{\gamma,\psi}(\pi')- V_{\gamma,\psi}(\pi'')}{V_{\gamma, \psi}(\pi)} dm_\pi(\gamma, \psi)
+ \int_{\gamma=1}  \left(D_{\gamma,\psi}(\pi')- D_{\gamma,\psi}(\pi'')\right) dm_\pi(\gamma, \psi).
\end{equation}
Similarly, by dividing (\ref{eq:M2}) by $\varepsilon^2$ and taking  the limit $\varepsilon\to 0$, we obtain
\begin{equation}\label{eq:ind2'}
0\geq \int_{\gamma<1} \frac{-1}{\alpha_{\rm max}^{\gamma,\psi}-1} \left(\frac{V_{\gamma,\psi}(\pi')- V_{\gamma,\psi}(\pi'')}{V_{\gamma, \psi}(\pi)}\right)^2 dm_\pi(\gamma, \psi).
\end{equation}

We now use (\ref{eq:ind1'})-(\ref{eq:ind2'}) to derive a couple of preliminary claims. 

\begin{claim}
For any $\pi$, $\int_{\gamma=1} D_{\gamma,\psi}(\cdot)dm_\pi(\gamma, \psi)$ is either ordinally equivalent to $C$ or 0. 

\end{claim}
\begin{proof}
We assume without loss that $C$ is non-constant, as otherwise the desired statement is immediate. Suppose the desired  statement does not hold, i.e.,  $ \int_{\gamma=1}  \left(D_{\gamma,\psi}(\pi')- D_{\gamma,\psi}(\pi'')\right) dm_\pi(\gamma, \psi)\not=0$ holds for some $\pi',\pi''$ with $C(\pi')=C(\pi'')$. Since $C$ is non-constant, we can assume $C(\pi')=C(\pi'')>0$. By abusing notations, let $\pi'^{\otimes k}$ and $\pi''^{\otimes k}$ denote the distribution of posteriors induced by $\mu'^{\otimes k}$ and $\mu''^{\otimes k}$ respectively, where $\mu'$ and $\mu''$ are experiments that induce $\pi'$ and $\pi''$ respectively. Then we have $C(\pi'^{\otimes k})=C(\pi''^{\otimes k})$ and $D_{\gamma,\psi}(\pi'^{\otimes k})-D_{\gamma,\psi}(\pi''^{\otimes k})=k(D_{\gamma,\psi}(\pi')-D_{\gamma,\psi}(\pi''))$ for all $k$. At the same time, both $V_{\gamma,\psi}(\pi'^{\otimes k})=(V_{\gamma,\psi}(\pi'))^{k}$ and $V_{\gamma,\psi}(\pi''^{\otimes k})=(V_{\gamma,\psi}(\pi''))^{k}$ converge to 0 as $k\to\infty$ at $\gamma<1$. Thus RHS of (\ref{eq:ind1'}) evaluated at $\pi'^{\otimes k}$ and $\pi''^{\otimes k}$ diverges as $k\to\infty$, which is a contradiction. 
\end{proof}

\begin{claim}
For any $\pi$  and  $(\gamma,\psi)$ in the support of $m_\pi$ with $\gamma<1$,  $V_{\gamma,\psi}$ is ordinally equivalent to $C$. 
\end{claim}
\begin{proof}
By (\ref{eq:ind2'}), for any $(\gamma,\psi)$ in the support of $m_\pi$ with $\gamma<1$,  $V_{\gamma,\psi}(\pi')=V_{\gamma,\psi}(\pi'')$. Since this holds for all $\pi',\pi''$ such that $C(\pi')=C(\pi'')$, $V_{\gamma,\psi}$ is ordinally equivalent to either $C$ or $-C$. The latter possibility is ruled out, because $C$ is non-constant and both  $V_{\gamma,\psi}$ and $C$ are Blackwell monotone.
\end{proof}

\begin{claim}

\begin{enumerate} 
\item Take any $(\gamma,\psi), (\gamma', \psi')$ with $\gamma, \gamma'<1$, such that $D_{\gamma,\psi}$ and $D_{\gamma',\psi'}$ are ordinally equivalent. Then $(\gamma,\psi)=(\gamma', \psi')$.


\item Take any $m,m'$ such that $\int_{\gamma=1} D_{\gamma,\psi}(\cdot)dm(\gamma, \psi)$ and $\int_{\gamma=1} D_{\gamma,\psi}(\cdot)dm'(\gamma, \psi)$ are ordinally equivalent. Then  
\[
\int_{\gamma=1} D_{\gamma,\psi}(\mu)dm(\gamma, \psi)=\sum_{i,j}\beta_{ij}{\rm KL}(\mu_i\|\mu_j), \;\;
\int_{\gamma=1} D_{\gamma,\psi}(\mu)dm'(\gamma, \psi)=\lambda\sum_{i,j}\beta_{ij}{\rm KL}(\mu_i\|\mu_j)
\] for some $\beta=(\beta_{i,j})_{i,j\in\Theta}$ and $\lambda>0$.
\end{enumerate}

\end{claim}
\begin{proof}

\noindent{\bf First part:} Suppose toward a contradiction that $(\gamma,\psi)\not=(\gamma', \psi')$. Then $\alpha^{\gamma,\psi}\not=\alpha^{\gamma',\psi'}$. Thus there exist $\beta,\beta'\in\mathbb R^\Theta_{++}$ such that $\alpha^{\gamma,\psi}\cdot\beta> \alpha^{\gamma,\psi}\cdot\beta'$ and $\alpha^{\gamma',\psi'}\cdot\beta<\alpha^{\gamma',\psi'}\cdot\beta'$.  Take experiments $\mu,\mu'$ with signal space $S=\{s_i: i\in \Theta\}\cup \{s^*\}$. The signal probabilities at each $i\in\Theta$ is given by
\[
\mu_i(s^*)=\exp[-\beta_it], \; \mu_i(s_j)=\exp[-t^2] \text{ for } j\not=i, \;  \mu_i(s_i)=1-\exp[-\beta_it]-(|\Theta|-1)\exp[-t^2],
\] 
\[
\mu'_i(s^*)=\exp[-\beta'_it], \; \mu'_i(s_j)=\exp[-t^2] \text{ for } j\not=i, \;  \mu'_i(s_i)=1-\exp[-\beta_it]-(|\Theta|-1)\exp[-t^2]
\] 
for some $t>0$.  As $t\to\infty$, both $V_{\gamma,\psi}(\mu)$ and $V_{\gamma,\psi}(\mu')$ vanish exponentially:  
\[
V_{\gamma,\psi}(\mu)=\exp[-t \alpha^{\gamma,\psi}\cdot \beta +o(t)], \;\; V_{\gamma,\psi}(\mu')=\exp[-t \alpha^{\gamma,\psi}\cdot \beta' +o(t)]
\]
where each $o(t)$ is such that $\frac{o(t)}{t}\to 0$. Thus $D_{\gamma,\psi}(\mu)>D_{\gamma,\psi}(\mu')$ for all large enough $t$. An analogous argument implies  $D_{\gamma',\psi'}(\mu)<D_{\gamma',\psi'}(\mu')$ for all large enough $t$, which is a contradiction.  

\noindent{\bf Second part:} By definition, 
\[
\int_{\gamma=1} D_{\gamma,\psi}(\mu)dm(\gamma, \psi)=\sum_{i,j}\beta_{ij}{\rm KL}(\mu_i\|\mu_j), \;\; 
\int_{\gamma=1} D_{\gamma,\psi}(\mu)dm'(\gamma, \psi)=\sum_{i,j}\beta'_{ij}{\rm KL}(\mu_i\|\mu_j)
\] for some $\beta=(\beta_{i,j})_{i,j\in\Theta}, \beta'=(\beta'_{i,j})_{i,j\in\Theta}\geq 0$. 
Take an experiment $\mu^{(ij)}$ for each distinct states $i,j$ such that ${\rm KL}(\mu^{(ij)}_i\|\mu^{(ij)}_j)=1$ and $\mu_k\not\in \{\mu^{(ij)}_i,\mu^{(ij)}_j\}$ is independent across  all $k\not=i,j$.  Then consider the mixture $\mu=\sum_{i,j}b_{ij} \mu^{(ij)}$ where $b=(b_{ij})_{i,j}\geq 0$.  Then $\sum_{i,j}\beta_{ij}{\rm KL}(\mu_i\|\mu_j)=\sum_{i,j}\beta_{ij}b_{ij}$ and $\sum_{i,j}\beta'_{ij}{\rm KL}(\mu_i\|\mu_j)=\sum_{i,j}\beta'_{ij}b_{ij}$. Thus for $\int_{\gamma=1} D_{\gamma,\psi}(\cdot)dm(\gamma, \psi)$ and $\int_{\gamma=1} D_{\gamma,\psi}(\cdot)dm'(\gamma, \psi)$ to be ordinally equivalent across these experiments, $\beta$ and $\beta'$ need to be proportional to each other.

\end{proof}

We now complete the proof of the second part of Theorem~\ref{prop:d-linear}.  It suffices to focus on the case in which $C$ is non-constant, as otherwise the desired statement is trivial.  First, consider the case  $m_\pi(\{1\}\times\Psi)>0$ for some $\pi$. Then by Claim 1, for every $\pi'$, $\int_{\gamma=1} D_{\gamma,\psi}(\pi')dm_{\pi'}(\gamma, \psi)$ is ordinally equivalent to $C$.  By Claim 3, there exist $\beta=(\beta_{ij})_{i,j\in\Theta}\geq 0$ and $\lambda_\pi>0$ for each $\pi$ such that 
\[
\int_{\gamma=1} D_{\gamma,\psi}(\cdot)dm_{\pi}(\gamma, \psi)=\lambda_\pi\sum_{i,j}\beta_{ij}{\rm KL}(\mu_i\|\mu_j).
\]
By Claim 2, for any $\pi$ and any $(\gamma, \psi)$ supported on $m_\pi$ with $\gamma<1$ is such that $D_{\gamma,\psi}$ is ordinally equivalent to $\sum_{i,j}\beta_{ij}{\rm KL}(\mu_i\|\mu_j)$. But this is not possible, because 
\[
\sum_{i,j}\beta_{ij}{\rm KL}(\mu_i\|\mu_j)=\sum_{i,j}\beta_{ij}{\rm KL}(\mu'_i\|\mu'_j), \;\; D_{\gamma,\psi}(\mu)>D_{\gamma,\psi}(\mu')
\] 
for $\mu'=\frac{1}{2}\mu^{\otimes 2}+\frac{1}{2}\nu$ in which $\mu$ is non-degenerate and $\nu$ is uninformative. Therefore  for every $\pi$ ,  $m_\pi([\frac{1}{|\Theta|}, 1) \times\Psi)=0$, and thus $C(\pi)=\lambda_\pi\sum_{i,j}\beta_{ij}{\rm KL}(\mu_i\|\mu_j)$.  Since $C(\pi)=\max_{\pi'}\lambda_{\pi'}\sum_{i,j}\beta_{ij}{\rm KL}(\mu_i\|\mu_j)$ for each $\pi$, we need to have $\lambda_\pi=\lambda_{\pi'}$ for all $\pi,\pi'$ that are not uninformative. This shows that $C$ is a KL cost.

Next, consider the case   $m_\pi(\{1\}\times\Psi)=0$ for all $\pi$. Since $C$ is non-constant, $m_\pi([\frac{1}{|\Theta|}, 1) \times\Psi)>0$ for all $\pi$.  By Claims 2-3,  $m_\pi$ is supported on value $(\gamma,  \psi)$ with $\gamma<1$ that is common across all $\pi$. Thus  $C(\pi)=m_\pi (\{\gamma, \psi\}) D_{\gamma, \psi}(\pi)$ for each $\pi$. Since $C(\pi)=\max_{\pi'}m_{\pi'} (\{\gamma, \psi\}) D_{\gamma, \psi}(\pi)$ for all $\pi$, we need to have $m_{\pi} (\{\gamma, \psi\})=m_{\pi'} (\{\gamma, \psi\})$ for all $\pi,\pi'$ that are not uninformative. This shows that $C$ is proportional to $D_{\gamma,\psi}$, as desired.

\subsection{Proof of Claim~\ref{cl:ex}}

\subsubsection{First part} 

We consider more general cost function of the form $C(\mu)=\lambda(R_{t}(\mu_0\| \mu_1)+R_{t}(\mu_1\| \mu_0))$ for some $t\in (0,1)$ and $\lambda>0$. This is a symmetric R\'{e}nyi cost if $t=1/2$.  The problem is written as
\[ \max_{\mu} \;  \sum_{i=0,1}\left[ \frac{1}{2} \left(  v  \mu_i (i) + w  \mu_i (\phi) \right) - \dfrac{\lambda}{t-1} \log \left(  \sum_{j \in A}  \mu_i (j)^t  \mu_{1-i} (j)^{1-t}  \right)\right]  . \]
The following shows that every optimal policy invokes three actions under certain parameters.
  

We first restrict attention to symmetric experiments $\mu$, i.e.,  $\mu_0 (0)= \mu_1 (1)$ and $\mu_0 (\phi)=\mu_1 (\phi)$ hold. Such experiments can be described by two numbers $\alpha,\pi\in [0,1]$ where $\alpha=1-\mu_0(\phi)=1-\mu_1(\phi)$ and $\alpha \pi=\mu_0(0)=\mu_1(1)$. The problem becomes  $\max_{\alpha, \pi\in [0,1]} V(\alpha, \pi)$, where the objective
\[
 V(\alpha, \pi):= v \alpha \pi +  w (1-\alpha) +  \frac{\lambda}{1-t} \log \left(  (1-\alpha)+ \alpha \pi^t (1-\pi)^{1-t}  + \alpha (1-\pi)^t \pi^{1-t}      \right)  \]
  is concave in $\pi$ and $\alpha$.  It holds that 
\begin{eqnarray*}
\frac{\partial V (\alpha, \pi)}{\partial \alpha}= v \pi - w +  \frac{\lambda}{1-t} \frac{ \pi^t (1-\pi)^{1-t} +  (1-\pi)^t \pi^{1-t}  -1    }{(1-\alpha) +\alpha \pi^t (1-\pi)^{1-t} + \alpha (1-\pi)^t \pi^{1-t}     } \\
\frac{\partial V (\alpha, \pi)}{\partial \pi}=  v \alpha +  \frac{\alpha\lambda}{1-t} \frac{ t (\frac{1-\pi}{\pi})^{1-t} + (1-t)  (\frac{1-\pi}{\pi})^t - (1-t) (\frac{\pi}{1-\pi})^t - t (\frac{\pi}{1-\pi})^{1-t}  }{(1-\alpha)+\alpha \pi^t (1-\pi)^{1-t}  + \alpha (1-\pi)^t \pi^{1-t}.    }  
\end{eqnarray*}
Observe that under $\alpha>0$, $\frac{\partial V (\alpha, \pi)}{\partial \pi}$ goes to $-\infty$ (resp. $\infty$) as $\pi\to 1$ (resp. $\pi\to 0$).

We now fix any $K>\frac{\lambda}{1-t}$. 
Consider the equation $\frac{\partial V(1, \pi)}{\partial \pi}=0$ at $\alpha>0$, i.e., 
\begin{equation}\label{eq:pi-v}
 v  +  \frac{\lambda}{1-t} \frac{ t (\frac{1-\pi}{\pi})^{1-t} + (1-t)  (\frac{1-\pi}{\pi})^t - (1-t) (\frac{\pi}{1-\pi})^t - t (\frac{\pi}{1-\pi})^{1-t}   }{ \pi^t (1-\pi)^{1-t}  + (1-\pi)^t \pi^{1-t}     } =0.
\end{equation}
(\ref{eq:pi-v}) admits a unique solution, denoted by $\pi_v$, which converges to 1 as $v\to\infty$. Take $\overline v$ large so that   $\frac{\lambda}{1-t}\left(1- \frac{ 1    }{\pi^t_v (1-\pi_v)^{1-t} +  (1-\pi_v)^t \pi_v^{1-t} }\right)<-K$ under any $v\geq\overline v$.

Fix any $v\geq \overline v$ and $w\in (v\pi_v-K, v\pi_v-\frac{\lambda}{1-t})$.  Then 
\[
\frac{\partial V(1, \pi_v)}{\partial \alpha}=v \pi_v - w +  \frac{\lambda}{1-t}\left(1- \frac{ 1    }{\pi^t_v (1-\pi_v)^{1-t} +  (1-\pi_v)^t \pi^{1-t}_v }\right)<v \pi_v - w -K<0
\]
so that $\alpha=1$ is not optimal. In addition, 
   \[
\frac{\partial V(0,\pi_v)}{\partial \alpha}=   v\pi_v-w+\lambda\frac{\pi^t_v (1-\pi_v)^{1-t} +  (1-\pi_v)^t \pi^{1-t}_v  -1 }{1-t}>v\pi_v-w-\frac{ 1 }{1-t} >0.
   \]
Note $\max_{\pi\in [0,1]}V(0,\pi)=V(0,\pi_v)$, since $V(0, \pi)$ is constant in $\pi$. Thus $\alpha=0$ is not optimal.  Thus $\alpha\in (0,1)$ and $\pi\in (0,1)$ hold at optimum, i.e., every action is chosen with a positive probability.

We now show more generally that any $\mu$ with ${\rm supp}(\mu)\not=A$ is suboptimal. 
We first observe that asymmetric experiments are dominated by symmetric ones. Under any $\mu$, take experiment $\tilde\mu$ such that $\tilde\mu_i (i) := \mu_{1-i} (1-i)$ and $\tilde\mu_i (\phi) := \mu_{1-i} (\phi)$ for $i=0,1$.  
Observe that 
\[
 \sum_{i=0,1} \frac{1}{2} \left(  v \left(\frac{1}{2}\mu_i(i)+\frac{1}{2}\tilde\mu_i(i)\right) + w  \left(\frac{1}{2}\mu_i(\phi)+\frac{1}{2}\tilde\mu_i(\phi)\right) \right) = \sum_{i=0,1} \frac{1}{2} \left(  v  \mu_i (i) + w  \mu_i (\phi) \right).\]
Moreover, we have that
\begin{equation*}\label{eq:convex reduction}
 R_t \left( \frac{1}{2} \mu_1 + \frac{1}{2} \tilde\mu_1\Big\| \frac{1}{2} \mu_0 + \frac{1}{2} \tilde\mu_0\right)=R_t \left( \frac{1}{2} \mu_0 + \frac{1}{2} \tilde\mu_0 \Big\| \frac{1}{2} \mu_1 + \frac{1}{2} \tilde\mu_1\right) \leq \frac{1}{2} R_t (\mu_0\| \mu_1) + \frac{1}{2} R_t (\mu_1\| \mu_0), 
 \end{equation*}
where the inequality follows from convexity of R\'{e}nyi divergences and $R_t (\tilde\mu_0\| \tilde\mu_1) = R_t (\mu_1\| \mu_0)$. The inequality is strict if $\mu_0(a)\tilde\mu_1(a)\not=\mu_1(a)\tilde\mu_0(a)$ for some $a\in A$ \citep[][Theorem 11]{van2014}.  
 It follows that $C (\frac{1}{2}\mu+\frac{1}{2}\tilde\mu) \leq C(\mu)$, and thus $\frac{1}{2}\mu+\frac{1}{2}\tilde\mu$ does weakly better than $\mu$. 

If ${\rm supp}(\mu)$ is either $\{0\}, \{1\}, \{\phi\}$, or $\{0,1\}$ then ${\rm supp}(\frac{1}{2}\mu+\frac{1}{2}\tilde\mu)\not=A$, which implies that both $\frac{1}{2}\mu+\frac{1}{2}\tilde\mu$ and $\mu$ are suboptimal. Now consider the remaining case  ${\rm supp}(\mu)=\{i,\phi\}$ for some $i=0,1$.  
For $\mu$ to be optimal, by the above argument we need  $\mu_0(a)\tilde\mu_1(a)=\mu_1(a)\tilde\mu_0(a)$ for all $a\in A$. This equality for $a=\phi$ implies $\mu_0(\phi)=\mu_1(\phi)$, and hence $\mu_0(i)=\mu_1(i)$. But this is suboptimal given $v>w$. 


\subsubsection{Second part}\label{app:ps}
%


\noindent{\bf Posterior separable cost}:
Fix any $w>0$. 
To simplify notation, we identify DM's belief as the probability of state 1. Let $U_v(p):=\max\{vp, v(1-p), w  \}-H(p)$ for each $p\in [0,1]$.

Suppose toward a contradiction that there are two values $v'<v''$ and corresponding optimal policies $\mu_{v'}, \mu_{v''}$ such that  ${\rm supp}(\mu_{v'})={\rm supp}(\mu_{v''})=A$.  For each $v=v',v''$,  let $p^i_v$ denote the posterior induced by $\mu_v$ at which action $i \in \{0,1, \phi\}$ is chosen.

By $v''>v'$, we have $p^\phi_{v''}\in (p^0_{v'}, p^1_{v'})$, which ensures $\alpha p^0_{v'} + (1- \alpha) p^1_{v'} = p^\phi_{v''}$ for some $\alpha\in (0,1)$. Then 
\begin{eqnarray*}
    \alpha U_{v''} (p^0_{v'}) + (1-\alpha) U_{v''} (p^1_{v'}) & > &  \alpha U_{v'} (p^0_{v'}) + (1-\alpha) U_{v'} (p^1_{v'}) \\
    & \geq & U_{v'} (p^\phi_{v''}) =U_{v''} (p^\phi_{v''})  \; ,
\end{eqnarray*}
where the first inequality uses $v''>v'$ and the second uses the optimality of $\mu_{v'}$ via the splitting argument \citep[e.g.,][]{gentzkow2014}. 
At the same time, the optimality of $\mu_{v''}$ implies $\alpha U_{v''} (p^0_{v'}) + (1-\alpha) U_{v''} (p^1_{v'})\leq U_{v''} (p^\phi_{v''})$, which is a contradiction. 


\

\noindent{\bf Max-KL cost}:  Let $V^*$ denote the highest value that DM can achieve when she restricts attention to policies $\mu$ such that $\mu_0(0)=\mu_1(1)=\pi$ and $\mu_0(1)=\mu_1(0)=1-\pi$ for some $\pi\in [0,1]$.  We focus on the case $w\not=V^*$ and show that any policy $\mu$ with ${\rm supp}(\mu)=A$  is sub-optimal. This implies the desired claim, as $V^*$ is strictly increasing in $v$. As in the proof of the first part, DM's value under $\mu$ is weakly lower than the value symmetric policy as given by $\frac{1}{2}\mu+\frac{1}{2}\tilde\mu$.  Thus, it suffices to assume that $\mu$ is symmetric.  

Given the symmetry, $\mu$ can be written as a convex combination of the form $\mu=\alpha \mu^\pi+(1-\alpha)\nu$ for some $\alpha\in (0,1)$ and $\pi\in [0,1]$, where $\mu^\pi_0(0)=\mu^\pi_1(1)=\pi$, $\mu^\pi_0(1)=\mu^\pi_1(0)=1-\pi$, and $\nu_0(\phi)=\nu_1(\phi)=1$.    
By dilution linearity, DM's value under policy $\alpha \mu^\pi+(1-\alpha)\nu$ equals to $\alpha V(\pi)+(1-\alpha)w$, where $V(\pi)$ is DM's value under policy $\mu^\pi$.  
If $w>V^*$ then DM's value under $\mu$ is strictly lower than $w$, which can be achieved by policy $\nu$. If $w<V^*$, DM's value under $\mu$  is strictly lower than $V^*$, which can be achieved by policy $\mu^{\pi^*}$ for some $\pi^*$.

\subsection{Proof of Proposition~\ref{thm:pairwise}}
The proof argument is analogous to those for Theorem~\ref{thm:subadditive} and Theorem~\ref{thm:general}, so we only provide a sketch.   
First, we extend Rényi divergence by allowing for parameter $t\in (1, \infty]$. The case of $t\in (1, \infty)$ is defined as in the main text, while the case of $t=\infty$ is defined by 
\[
R_\infty(\mu_i\|\mu_j):=\log \; {\rm esssup}_{s \in S} \frac{d\mu_i}{d\mu_j} (s).
\]

Let $\Psi':=\cup_{i,j\in\Theta, i\not=j}\{e_i-e_j\}\subseteq\Psi$ be the set of $\psi$s that are non-zero only at binary coordinates. 
Then observe for any $\gamma\in [\frac{1}{2}, \infty]$ and $\psi=e_i-e_j$, $D_{\gamma,\psi}(\mu)=R_t(\mu_i\| \mu_j)$ for $t=\gamma$. 

As in Lemma~\ref{lem:monotone}, we represent the cost function as a monotone functional of divergences restricted to $[\frac{1}{2}, 1]\times\Psi'$. To do so, instead of using Blackwell monotonicity to apply Theorem 19 in \cite{farooq2024},  we use pairwise Blackwell monotonicity to apply Theorem 1 in \cite{mu2021}. That is, if $R_t(\mu_i\|\mu_j)>R_t(\nu_i\|\nu_j)$ for each pair of states $i,j$ and $t\geq \frac{1}{2}$, then $\mu^{\otimes k}$ pairwise Blackwell dominates $\nu^{\otimes k}$ for some $k$. 

By following the same argument as in the proof of Theorem~\ref{thm:general}, we obtain $C$ is represented by Max-Rényi cost in which each measure $m$ is supported on $[\frac{1}{2}, \infty]\times\Psi'$. Finally, by following the same argument as in the proof of Theorem~\ref{thm:subadditive}, we conclude that each measure $m$ is supported on $[\frac{1}{2}, 1]\times\Psi'$, which yields a representation of the desired form.




\section{Discussions}
\subsection{Sub-additive Posterior-Separable Cost}\label{app:UPS}
Consider a posterior-separable cost function $C_q(\mu)=\int (H(q)-H(p)) d\pi^\mu_q(p)$, making its dependence on the prior $q$ explicit. 
We allow for the prior $q$ to vary--maintaining full support--while keeping $H$ fixed, i.e., the cost is uniformly posterior separable. 

%
%

Below we consider binary states $\Theta=\{0,1\}$ for simplicity. We identify each belief $p$ with the probability on state 1.  
As shown in Appendix A.1 of \cite{bloedel2020}, $C_q$ satisfies sub-additivity at all $q$ if and only if $F (p):= p^2 (1-p)^2 H^{\prime \prime} (p)$ is convex.

For example, under Shannon entropy 
\[
H(p)= - p \log (p) - (1-p) \log (1-p),
\] we have $F (p)= - p (1-p)$, which is strictly convex.
 Therefore this cost function is sub-additive, as was shown by \cite{lindley1956}.

Let us now consider the generalized entropy of \cite{tsallis1988} given by 
\[
H(p)= \frac{1}{\sigma-1} (1 - p^\sigma - (1-p)^\sigma),\] 
which is parametrized by $\sigma>0, \sigma \neq 1$. This cost function is used for example by \cite{caplin2022}. Shannon entropy is the extreme case as $\sigma \rightarrow 1$. 
Here we have $F(p)= - \sigma (p^\sigma (1-p)^2 + p^2 (1-p)^\sigma)$.  
It turns out that sub-additivity fails at $\sigma>1$, no matter how close to 1, even though 
sub-additivity holds strictly under Shannon entropy.  
To see this, by denoting $x:= \frac{p}{1-p}$, $F^{\prime \prime} (p) \geq 0$ can be equivalently written as
\begin{equation*}\label{eq:xsigma}
    2 (1 + x^\sigma) + \sigma (\sigma-1) (x^2 + x^{\sigma-2}) - 4 \sigma (x^{\sigma-1} + x) \leq 0 \; .   
\end{equation*}
For $\sigma>1$, the LHS  goes to $+ \infty$ as $x \rightarrow \infty$ ($p \rightarrow 1$), so that sub-additivity is violated.

%

\subsection{Other Information Measures}\label{app:others}

Max-Rényi cost can be related to information measures used in broader contexts. 
Below we provide two examples, focusing on binary states $\Theta=\{0, 1\}$ for simplicity. 

First, Chernoff information is often used to quantify the value of information in learning settings: \citep[e.g., ][]{moscarini2002}.  
This takes the form
\[
-\min_{t\in [-1, 1]} \log \int \left(\frac{d\mu_0}{d\mu_1} (s)  \right)^{t-1} d\mu_0(s)=\max_{t\in [-1, 1], i\in\{0,1\}} (1-t) R_t(\mu_i\| \mu_{1-i}).
\]
Therefore it is a special case of Max-Rényi cost.   

Second,  the following measure is often used to quantify the amount of information in the context of consumer privacy \citep{dwork2006}: 
\[
{\rm esssup}_{s}\left| \log \frac{d\mu_1}{d\mu_0}(s) \right|=\max\{ R_\infty(\mu_0\|\mu_1), R_{\infty}(\mu_1\|\mu_0)\}.
\]
This it is an instance of the generalized Max-Rényi cost in Theorem~\ref{thm:general}, in which each measure $m$ is such that $m(\cdot, \Psi)$ is supported on $\infty$. This function does not satisfy dilution convexity. Indeed, as is shown below,  this cost is  \textit{\textbf{maximally dilution concave}:} $C(\alpha \mu+(1-\alpha)\phi)=C(\mu)$ for all $\mu$, uninformative $\phi$, and $\alpha\in (0,1)$.

\begin{proposition}\label{prop:cs}
Assume $\Theta=\{0,1\}$.  $C$ is Blackwell monotone, maximally dilution concave, sub-additive, and identity additive if and only if there is a closed convex set $B\subseteq \mathbb R^2_+$ such that
\[
C(\mu)=\max_{(\beta_{0}, \beta_{1})\in B}\beta_{0}R_\infty(\mu_0\| \mu_1)+\beta_{1}R_\infty(\mu_1\| \mu_0).
\]
\end{proposition}

\begin{proof}
Observe that ``if'' direction follows from $R_\infty(\alpha\mu_i+(1-\alpha)\phi_i\| \alpha\mu_j+(1-\alpha)\phi_j)=R_\infty(\mu_i\| \mu_j)$ for any $\alpha\in (0,1)$. 
To establish ``only if,'' under any Max-Rényi cost that is maximally dilution concave,  
\begin{eqnarray*}
C(\mu) &=& C (\alpha \mu+(1-\alpha)\phi) \\
&=&\max_{m\in M} \sum_{i,j\in \Theta}  \int_{\mathbb R_{++}} R_t(\alpha\mu_i+(1-\alpha)\phi\|\alpha \mu_j+(1-\alpha)\phi_j) dm_{ij}(t)+ m_{ij}(\{\infty\})R_\infty(\mu_i\| \mu_j).
\end{eqnarray*}

By taking  $\alpha \searrow 0$, $C(\mu) = \max_{m\in M} \sum_{i,j\in \Theta}  m_{ij}(\{\infty\})R_\infty(\mu_i\|\mu_j)$. Thus the representation obtains with $B:=\{ (m_{01}(\{\infty\}), m_{10}(\{\infty\})) : m\in M \}$.
\end{proof}
\footnotesize
\begin{spacing}{0.05}
\bibliographystyle{econometrica}
\bibliography{infocost}
\end{spacing}

\end{document}